\documentclass[fleqn,usenatbib]{mnras}

\usepackage{newtxtext,newtxmath}

\usepackage[T1]{fontenc}
\usepackage{ae,aecompl}

\usepackage{graphicx}	
\usepackage{amsmath}	
\usepackage{amssymb}	
\usepackage{array}

\newcommand{\dix}[1]{\cdot 10^{#1}}
\newcommand{\dif}[1]{\mathrm{d}#1} 
\newcommand{\deriv}[2]{\frac{\mathrm{d}#1}{\mathrm{d}#2}} 
\newcommand{\pderiv}[2]{\frac{\partial #1}{\partial #2}} 
\newcommand{\norm}[1]{\ensuremath{\left\Vert #1 \right\Vert}} 

\newcommand{\Msol}{M_{\odot}}

\title[Spider timing model]{A spider timing model: accounting for quadrupole deformations and relativity in close pulsar binaries}

\author[G. Voisin et al.]{
Guillaume Voisin$^{1,2}$\thanks{E-mail: guillaume.voisin@manchester.ac.uk; astro.guillaume.voisin@gmail.com},
Ren\'e P. Breton$^{1}$,
Charlotte Summers$^{1,3}$
\\
$^{1}$Jodrell Bank Centre for Astrophysics, School of Physics and Astronomy, The University of Manchester, Manchester M19 9PL, UK\\
$^{2}$LUTH, Observatoire de Paris, PSL Research University, 5 Place Jules Janssen, 92195 Meudon, France\\
$^{3}$ School of Mathematics \& Maxwell Institute, University of Edinburgh, Edinburgh, EH9 3FD, UK
}

\date{Accepted XXX. Received YYY; in original form ZZZ}

\pubyear{2018}

\begin{document}
\label{firstpage}
\pagerange{\pageref{firstpage}--\pageref{lastpage}}
\maketitle

\begin{abstract}
Spider millisecond pulsars are, along with some eclipsing post-common envelope systems and cataclysmic variables, part of an expanding category of compact binaries with low-mass companions for which puzzling timing anomalies have been observed. The most prominent type of irregularities seen in them are orbital period variations, a phenomenon which has been proposed to originate from changes in the gravitational quadrupole moment of the companion star. A physically sound modelling of the timing of these systems is key to understanding their structure and evolution.
In this paper we argue that a complete timing model must account for relativistic corrections as well as rotationally and tidally induced quadrupole distortions. We solve for the resulting orbital dynamics using perturbation theory and derive the corresponding timing model in the low eccentricity limit.
We find that the expected strong quadrupole deformation of the companion star results in an effective minimum orbital eccentricity. It is accompanied by a fast periastron precession which, if not taken into account, averages out any measurement of the said eccentricity. We show that, with our model, detection of both eccentricity and precession is likely to be made in many if not all spider pulsar systems. Combined with optical light curves, this will allow us to measure the apsidal motion constant, connecting the quadrupole deformation to the internal structure, and thus opening a new window into probing the nature of their exotic stellar interiors. Moreover, more accurate timing may eventually lead spider pulsars to be used for high-precision timing experiments such as pulsar timing arrays.

\end{abstract}




\section{Introduction}
Millisecond pulsars (MSPs) are thought to be the offspring of low-mass X-ray binaries (LMXBs) in which mass transfer from a low-mass star onto the neutron star carries angular momentum and spins it up to a very fast rotation \citep{alpar_new_1982}. This `pulsar recycling' process is evidenced from the fact that a large fraction of the MSP population lies in binary systems with evolved companions and that the mass of these neutron stars tends to be larger that one of pulsars found in binary system that have experienced reduced mass transfer episodes \citep{ozel_masses_2016, antoniadis_millisecond_2016, tauris_formation_2017}. More recently, the link between LMXBs and MSPs became clearer as some systems, now commonly referred to as transitional MSPs (tMSPs) have been observed to transition from one class to the other \citep{archibald_radio_2009, papitto_swings_2013, bassa_state_2014}.

An important growing class of binary MSPs is collectively nicknamed as spiders due to the fact that the energetic pulsar strongly irradiates (and partially destroys) their companion. Contrary to most MSPs that have white dwarf companions, spiders are found in tight, sub-day orbits, and are believed to be linked to the evolutionary path of so-called converging systems \citep{chen_formation_2013}. The two primary sub-types of spiders are identified as black widows and redbacks according to the mass of the secondary \citep[$m_c \lesssim 0.05\Msol$ and $m_c\gtrsim 0.2\Msol$, respectively][]{roberts_surrounded_2012}, with the latter one linked tMSPs. There is tantalising evidence that some spider pulsars lie at the higher end of the neutron star mass range \citep{van_kerkwijk_evidence_2011,linares_peering_2018}, which may indicate that a key factor in their evolution is responsible for it.

A disproportionately large fraction of the MSP population belongs to the spider category. Notable examples are the two fastest known spinning pulsar, PSR~J1748-2446ad \citep{hessels_radio_2006} and PSR~J0952-0607 \citep{bassa_lofar_2017}, which are a redback and a black widow, respectively. While in theory these spiders should contribute a large number of sources to be used for pulsar timing arrays \citep[see, e.g.,][]{manchester_millisecond_2017}, their timing displays undesirable behaviours which greatly hinders their usefulness as reliable standard clocks. A better modelling of these phenomena may therefore allow for a significant improvement of their timing solution.

Perhaps the most common limitation to the timing of spiders are the prominent, sometimes long lasting radio eclipses caused by outflowing material driven off the surface of the companion by the pulsar wind \citep[see, e.g.,][and references therein for recent work on the topic]{polzin_low-frequency_2018}. These eclipses not only cause a disappearance of the pulsar at radio frequencies but also chromatic effects affecting the pulse such as dispersion delays and scattering (\citet{stappers_nature_2001}, Polzin in prep.). As a consequence, radio eclipses severely limit our ability to detect relativistic effects such as the Shapiro delay, and thus they deprive us from additional constraints on system parameters such as masses and orbital inclination \citep[see, e.g.,][and references therein]{edwards_tempo2_2006}. However, the fact that the eclipsing material is transparent to high-energy photons, makes their timing in the gamma rays (with FERMI/LAT) an promising venue to consider \citep{pletsch_gamma-ray_2015}.

Due to their evolution, the companion star in spider systems should be tidally locked and their orbit circularised \citep[e.g.][]{hurley_evolution_2002}. There is some evidence for a very small (potentially superficial) asynchronous rotation in PSR~J1723-2837 \citep{van_staden_active_2016}, but apart from this exceptional case, one might be tempted to think that the timing model for spider pulsars should in principle be very simple. Surprisingly, some spider systems such PSRs~J2051-0827 and J1731-1847 display small, but tangible eccentricity \citep{lazaridis_evidence_2011, ng_high_2014}. While this puzzling characteristic should not prevent an accurate timing precision to be achieved, it appears that a large fraction of the spiders also suffer from important orbital period variations. For instance, the original black widow PSR~B1957+20 \citep{fruchter_millisecond_1988} shows orbital period variations having an amplitude $\Delta P / P \sim 10^{-7}$ over a time scale of 5 to 10 years \citep{arzoumanian_orbital_1994}. Since then several other systems have shown similar variations (see Section \ref{sec:application}), and it seems plausible that it is a common feature to all spider pulsars. In current timing models \citep[see, e.g.,][]{pletsch_gamma-ray_2015,shaifullah_21_2016}, orbital period variations are empirically fitted using a series of orbital period derivatives.

Around the time when PSR~B1957+20 was discovered, mechanisms were proposed in order to explain pseudo-periodic variations of orbital periods in various types of close-binary systems such as Algol and cataclysmic variables \citep{applegate_mechanism_1992}, and later extended to include PSR~B1957+20 \citep{applegate_orbital_1994}. The main idea the Applegate mechanism is that magnetic cycles observed in magnetically active stars may also produce quasi-periodic variations of the gravitational quadrupole moment of the star through the interplay of magnetic pressure and/or tension. This mechanism also predicts that luminosity and orbital variations should correlate which, to the best of our knowledge, is yet to be proven in the case of spiders. Other authors \citep{lanza_orbital_1998} proposed that luminosity variations might be much lower than first estimated \citet{applegate_mechanism_1992}, as the luminosity response might not be instantaneous but rather spread over a much longer Kelvin-Helmoltz time scale. While magnetic cycles may not be the underlying driving force, it is generally agreed that quadrupole variations offer an interesting interpretation for period variations, though it has failed so far to provide much predictive power. One advantage of this type of mechanism is to naturally explain the time scale of orbital period variations and without requiring orbital torques, as the induced force is purely radial.

It should also be noted that spider companions are considered exotic due to their irradiation, fast rotation and low mass (especially in the case of black widows). As a result, less is known about their internal structure than in the case of other stars, and in particular main sequence stars \citep{lanza_orbital_1998}. Several important questions therefore remain unanswered about their nature. It is for instance unclear as to whether redbacks and black widows are a single class of objects at different evolutionary stages or if they evolved from two separate populations \citep{benvenuto_evolutionary_2012,chen_formation_2013}. Thus, probing the internal structure of the companion stars would greatly help in understanding and constraining models of their past and future evolution. Additionally spider pulsar companions lie at the most extreme range of irradiated systems, beyond hot Jupiters and donor stars in cataclysmic variables \citep{hernandez_irradiated_2016}. The process through which their day side gets heated must considerably affect the structure and evolution of the star such as bloating them to the point of filling their Roche lobe \citet{podsiadlowski_irradiation-driven_1991}. It also appears that most of spider companions reprocess a rather universal fraction of the energy available from the pulsar spin-down \citep{breton_discovery_2013}. Another potential important piece of information that could be learn from a better understanding their structure and the mechanism which triggers tMSPs to experience episodes of active mass transfer followed by quiescent phases in which no disc is present.

In this work, we devise a timing model that encompasses and consistently extends the current state-of-art timing models for spiders and more generally any type of compact low-mass binary in which one of the two stars suffers from quadrupole distortion. In these systems, orbital period variations are assumed to be caused by a variable deformation of the star, as in \citet{applegate_orbital_1994}, however we do not need to make any assumption about the physical mechanism causing these deformations, which will allow to compare against different theories (see \ref{sec:tide}). This is possible thanks to the small magnitude of the observed orbital variations, which enables us to assume that feedback from orbital motion on internal structure is negligible. We therefore treat the effect of internal structure variations on orbital dynamics as given by an arbitrary function depending only on time, and focus on its implications for the timing of the system. Our model accounts for the relatively large quadrupole moment which spider companions are expected to bear due to their fast rotation and strong tidal interaction with the neutron star, which results in a well-known orbital precession. We demonstrate that the measurement of such a precession via timing can provide a direct insight into the structure of the companion star, thus potentially offering a way to answer some of the questions from above. Determining the precession rate of these systems relies on them having a measurable eccentricity, though we demonstrate that the perturbed orbits must necessarily have a minimum effective eccentricity that we expect to be detectable.

We also discuss and consistently include in our model relativistic corrections to first post-Newtonian order (1PN) as they are significant for the dynamics even if they are expected to be somewhat smaller in magnitude than the quadrupole interactions. The only relevant relativistic effect to be considered is also an orbital precession, which adds up to the precession from quadrupole origin. Therefore, we aim at providing a comprehensive timing model that can account simultaneously for relativistic and quadrupole effects (tidal and centrifugal, both static and time variable) based on first-order perturbation of Keplerian orbital dynamics of a binary systems which includes both secular and short-term contributions. To account for very low orbital eccentricities as observed in spider systems, we truncate our model to first order in eccentricity and use the Laplace-Lagrange parametrisation introduced in \citet{lange_precision_2001}. We take a particular care in making explicit the dynamical meaning of the model parameters. Although, the model is devised with spider systems in mind, it is clear that our results also apply to any low-mass, low-eccentricity binary pulsar.

In the remaining of this article, we first discuss the general effects of tidal interactions of the companion star in Section \ref{sec:tide}, solve the orbital dynamics to first order in relativistic and quadrupole perturbations in Section \ref{sec:motion}, which we use to obtain a timing model for spiders to first order in eccentricity derived in Section \ref{sec:timing}. In Section \ref{sec:application}, we estimate the orders of magnitude of the orbital variability and discuss the different effects in known spider systems as well as other systems. Finally, we give our conclusions in Section \ref{sec:conclusions}.

\newcommand{\xh}{\hat{x}}
\newcommand{\vxh}{\vec{\xh}}
\newcommand{\yh}{\hat{y}}
\newcommand{\vyh}{\vec{\yh}}
\newcommand{\zh}{\hat{z}}
\newcommand{\vzh}{\vec{\zh}}
\newcommand{\rh}{\hat{r}}
\newcommand{\vrh}{\vec{\rh}}
\newcommand{\vnh}{\vec{\hat{n}}}

\newcommand{\hQ}{\hat{Q}}
\newcommand{\qadim}{J_2}

\section{The effects of tidal interaction}
\label{sec:tide}

In this paper we consider spider evolution beyond the end of the low-mass x-ray binary phase. We assume that mass transfer is negligible and therefore that the main mechanism that can change orbital eccentricity and spin of the companion is tidal interaction (e.g. \citet{hurley_evolution_2002}). 

This assumption is not obvious, particularly in the case of transitional redbacks (\citet{kennedy_kepler_2018} and references therein). These systems alternate between X-ray binary and active radio pulsar phases on short time scales (years or less). However it seems that the amount of matter then transferred is rather small, as can be judged from their faint X-ray luminosity compared to other low-mass X-ray binaries (e.g. \citet{archibald_radio_2009}).
Generally, spider companions are known to evaporate. The fate of this matter is unclear, yet it seems unlikely that it be accreted by the neutron star, as the pulsar wind would tend to repel the material through the radio ejection mechanism \citep{burderi_radio_2002}. Besides, the rate of evaporation is uncertain, but has been shown to be fairly small in some cases \citep{polzin_low-frequency_2018}, where full evaporation is not expected within a timescale comparable with the Universe's age. In the following, we thus focus only on tidal interactions.

\subsection{Phase locking and circularisation}

The orbital separation is readily obtained from Kepler's third law, 
 \begin{equation}
 \label{eq:sep}
 a \simeq 3.5 \left(\frac{M}{1.7\Msol}\right)^{1/3} \left(\frac{P}{4\mathrm{h}}\right)^{2/3} \mathrm{ls},
 \end{equation}  
 where $M$ is the total mass of the system (which can be approximated to the mass of the neutron star), and $P$ is the orbital period.
 
 At small orbital separation such as is the case for spider binaries ($P \sim 1.5-24$\,hrs; equation \eqref{eq:sep}), tidal forces from the neutron star lead to tidal locking: the spin axis of the companion is aligned with the orbital angular momentum, and the spin period is equal to the orbital period and the orbital eccentricity tends to zero. We stress that we consider here the eccentricity of the osculating Keplerian orbit, as this will become important later. Tidal locking results from viscous dissipation of the tidal deformation of the companion star, a process that is most efficient in stars possessing important convective envelopes \citep{zahn_reprint_1977, hurley_evolution_2002, ogilvie_tidal_2014}. This is presumably more efficient in black widows which are thought to be fully convective stars (e.g. \citet{chen_formation_2013}), but probably also at the lower end of the redbacks mass range since stellar evolution sets the upper limit for fully convective stars at $0.2-0.3\Msol$ (e.g. \citet{rappaport_new_1983,podsiadlowski_evolutionary_2002,chen_formation_2013}). In the case of a fully convective star, the synchronization time scale is given by \citep{hut_tidal_1981, hurley_evolution_2002}
\begin{equation}
\label{eq:tausync}
	\tau_s = \frac{21}{7} \frac{\tau_\mathrm{conv}}{f_\mathrm{conv}} \frac{I_c}{m_cR_c^2} q^{-2} \left(\frac{a}{R_c}\right)^6,
\end{equation} 
where $\tau_\mathrm{conv}$ is the convection eddy turnover time scale, $f_\mathrm{conv}$ is a numerical factor that we will take equal to 1, $I_c, m_c, R_c$ are the moment of inertia, the mass and the radius of the companion star respectively (assumed spherically symmetric here), and $q=m_c/m_p$ is the ratio of the mass of the companion to the mass of the pulsar. In a fully convective star the convection time scale can be approximately expressed as \citep{hurley_evolution_2002}
\begin{equation}
	\tau_\mathrm{conv} = 0.4311 \left(\frac{m_cR_c^2}{6 L_c}\right)^{1/3},
\end{equation}
where $L_c$ is the luminosity of the star which we relate to the effective  temperature of the star by $L_c = 4\pi R_c^2 \sigma T_{\mathrm{eff}}^4$. Defining the filling factor of the companion star as the ratio between its volume-averaged radius and the volume-averaged radius of a Roche-lobe filling star $R_f$ we have $R_c = f R_f$. The volume-averaged Roche-lobe radius can be expressed in units of the orbital separation using Eggleton's approximation \citep{eggleton_aproximations_1983}, accurate to $1\%$, 
\begin{equation}
\label{eq:rhof}
	\rho_f(q) = \frac{R_f}{a} = \frac{0.49 q^{2/3}}{0.6q^{2/3} + \ln(1+ q^{1/3})} 
\end{equation}   
The synchronization time scale \eqref{eq:tausync} for fully convective stars is then given by 
\begin{eqnarray}
	\tau_s & = & 1.5\dix{6} \left(\frac{m_c}{0.1\Msol}\right)^{1/3} \left(\frac{T_{\mathrm{eff}}}{5000 \mathrm{K} }\right)^{-4/3} \left(\frac{I}{m_cR_c^2}\right) \\
	& & \left(\frac{q}{\frac{1}{16}}\right)^{-2} f^{-6}\left(\frac{\rho_f(q)}{\rho_f\left(\frac{1}{16}\right)}\right)^{-6} \mathrm{yr}. \nonumber
\end{eqnarray}
The above formula shows the strong dependence of the synchronization time scale on the mass ratio $q$. All else being equal, this time scale ranges from $\tau_s \sim  5\dix{3}$\,yr for $q \sim 0.3,\, m_c \sim 0.1\Msol$ (compatible with a redback system) to  $\tau_s \sim  6\dix{7}$\,yr for $q \sim 0.02,\, m_c \sim 0.03\Msol$ (compatible with a black widow system). In every case, this is significantly less than the time spent in the low-mass X-ray binary state according to the evolution model of \citet{chen_formation_2013}.

Similarly, the orbital circularisation time scale applied to a fully convective star is given by \citep{rasio_tidal_1996, hurley_evolution_2002},
\begin{eqnarray}
\tau_c & = & \frac{\tau_{\mathrm{conv}}}{f_{\mathrm{conv}}} q(1+q)\left(\frac{a}{R_c}\right)^{8} \\
\label{eq:circularisation}
\tau_c & \simeq &  8\dix{5}\left(\frac{m_c}{0.1\Msol}\right)^{1/3} \left(\frac{T_{\mathrm{eff}}}{5000 \mathrm{K} }\right)^{-4/3} \\
& & \frac{q(1+q)}{\frac{1}{16}\left(1 + \frac{1}{16}\right) } f^{-8}\left(\frac{\rho_f(q)}{\rho_f\left(\frac{1}{16}\right)}\right)^{-8} \mathrm{yr}, \nonumber
\end{eqnarray} 
with an even stronger dependence on $q$. The circularisation time scale ranges from $\tau_c \sim  6\dix{3}$\,yr for $q \sim 0.3,\, m_c \sim 0.1$ (compatible with a redback system) to  $\tau_c \sim  3\dix{7}$\,yr for $q \sim 0.02,\, m_c \sim 0.03$ (compatible with a black widow system). Additionally, we note that eccentricity is also suppressed as a result of mass transfer during the low-mass X-ray binary phase. 
  
Therefore we conclude that tidal interactions alone are sufficient to lock the companion in phase with the pulsar and circularise the orbit before the system even enters the spider state.

\subsection{The phase-locked quadrupole potential}
The frame of the principal axis of inertia (PAI)  $\left(\vxh,\vyh, \vzh\right)$ is the frame such that the star has 3 non-zero moments of inertia $A\leq B\leq C$ giving the inertia tensor 
\begin{equation}
\hat{I} = \left(\begin{matrix}
	A & 0 & 0 \\
	0 & B & 0 \\
	0 & 0 & C
\end{matrix}\right),
\end{equation}
with 
\begin{eqnarray}
	A = \int\dif{V} \rho(\vrh) (\yh^2 + \zh^2), \\
	B = \int\dif{V} \rho(\vrh) (\xh^2 + \yh^2), \\
	C = \int\dif{V} \rho(\vrh) (\zh^2 + \xh^2).
\end{eqnarray}
The quadrupole moment is related to the inertia tensor by 
\begin{equation}
\label{eq:Q}
	\hQ = \hat{I} - \frac{1}{3}\mathrm{Tr}( \hat{I}),
\end{equation}
where $Tr()$ takes the trace of a matrix. 

The gravitational potential due to a companion star bearing a quadrupole moment is a rotation-invariant quantity and thus can directly be expressed using PAI frame coordinates, 
\begin{equation}
\label{eq:potQ}
\Phi_c = - \frac{Gm_c}{\rh}- \frac{3}{2}\frac{G\vnh^T \hQ \vnh}{\rh^3},
\end{equation}
where $G$ is the gravitational constant, $m_c$ the mass the companion star, $\vnh = \vrh/\rh$ is the unit vector in the PAI frame of the direction going from the companion to the body undergoing the effect of the potential, here the pulsar, and $\vnh^T$ is the transposed vector. The first term of the above equation is the usual Newtonian monopolar potential, and the second term is the quadrupole potential.

In this paper, we assume that the companion star is tidally locked to the pulsar, such that the PAI frame vector $\vxh$ is aligned with the radial vector connecting the pulsar to the companion. This means that $\vrh \propto \vxh$ and therefore the numerator of the quadrupole potential in \eqref{eq:potQ} is merely the $\vxh\vxh$ component of the quadrupole matrix \eqref{eq:Q},  
\begin{equation}
\vnh^\bot \hQ \vnh = Q_{\xh\xh}= \frac{2}{3}A - \frac{1}{3}(B+C).
\end{equation}
In a two-body system with characteristic separation $a$ between the two bodies, the ratio between the quadrupole potential and the monopole potential is $\sim 3Q_{\xh\xh} / 2m_ca^2$ which suggests a natural way of defining a dimensionless quadrupole function,
\begin{equation}
\label{eq:qadim1}
	\qadim = \frac{3}{2}\frac{Q_{\xh\xh}}{m_ca^2},
\end{equation}
where $Q_{\xh\xh}$ is defined in equation \eqref{eq:Q}.

\newcommand{\vqadim}{J_v}
\newcommand{\bqadim}{\bar{\qadim}}
\newcommand{\rotqadim}{J_s}
\newcommand{\tidqadim}{J_t}
The quadrupole function contains contributions extrinsic and intrinsic to the star, the latter being independent of orbital motion. In the present case of a circularised and synchronized orbit, the spin-induced moment of the companion is an intrinsic contribution of constant magnitude $\rotqadim$. We also consider a variable intrinsic contribution $\vqadim(t)$. Although the physical origin of the latter need not be specified, it has been proposed that the stellar magnetic field applies a strain that may significantly distort the star and vary following magnetic cycles \citep{applegate_mechanism_1992, lanza_orbital_1998, lanza_orbital_1999}, resulting in observable orbital period variations, the so-called Applegate mechanism. In these simplified models, the variation of quadrupole momentum is related to variations of the angular momentum of a thin shell.  Refinements where later brought by \citet{brinkworth_detection_2006} who introduced a treatment based on a finite shell and a core, followed by \citet{volschow_eclipsing_2016} who generalised that approach to more realistic density profiles of the companion star. Alternatively, \citet{lanza_orbital_2005, lanza_time_2006} introduced a treatment of the complete continuous redistribution of angular momentum across the entire convective zone of the star, later expanded by \citet{volschow_physics_2018}. In all these models, the underlying magnetic cycle are imposed as ad hoc assumptions. Recently, the first 3D stellar simulation aimed at demonstrating self-consistently the connection between the magnetic dynamo and quadrupole moment variations has been performed by \citet{navarrete_magneto-hydrodynamical_nodate}. However qualitatively successful, this simulation is restricted in range and cannot assess fully realistic cases due to the overwhelming computing power needed. Most of the aforementioned models (in particular \citet{lanza_time_2006, volschow_eclipsing_2016, volschow_physics_2018, navarrete_magneto-hydrodynamical_nodate}) conclude that it is impossible to produce the observed magnitude of orbital period variations with the Applegate mechanism only, except maybe for post-common-envelope systems with companions slightly below the fully-convective mass limit $m_c \sim 0.35\Msol$ \citep{lanza_time_2006,volschow_physics_2018}. Interestingly, these might be akin to redback companions. In any case, the difficulty in modelling magnetic dynamo and its coupling to differential rotation, particularly in fast-rotating stars, leaves the question open to know if the Applegate mechanism is the cause of the observed variations. In our treatment, the variable component of the quadrupole, $\vqadim(t)$, can be specified arbitrarily (for example as a Taylor expansion, see \eqref{eq:nderivatives}) allowing to test virtually every model of quadrupole variations.

The extrinsic contribution to the quadrupole function results from the tidal field of the neutron star which contributes a term $\tidqadim \frac{a^3}{r^3}$ under our assumptions. The total quadrupole function then reads 
\begin{equation}
\label{eq:qdadimcomponents}
	\qadim = \rotqadim + \vqadim(t) + \tidqadim \frac{a^3}{r^3},
\end{equation}
and we can rewrite the potential of the companion, equation \eqref{eq:potQ}, as 
\begin{equation}
\label{eq:phic}
	\Phi_c = - \frac{Gm_c}{r} \left( 1 + \qadim(r,t)\frac{a^2}{r^2}\right).
\end{equation}

\subsubsection{Estimates of tidal-centrifugal deformations}
\label{sec:apsidalconstant}
The quadrupole components $\rotqadim$ and $\tidqadim$ of the companion are the leading order responses of its stellar structure to two external fields: the effective centrifugal potential arising from its spin, and the tidal field of the pulsar. The potential described by equation \eqref{eq:potQ} is then only valid outside the star. 

The coefficients $\rotqadim$ and $\tidqadim$ can be related to the deformation of the star through the apsidal motion constant $k_2$, such that \citep{sterne_apsidal_1939, kopal_dynamics_1978}
\begin{eqnarray}
\label{eq:rotqadim}
\rotqadim & = & - k_2 \frac{1}{3}\frac{R_c^2 n^2}{Gm_c/R_c}\left(\frac{R_c}{a}\right)^2, \\
\tidqadim & = & -  k_2 \frac{m_p}{m_c}\left(\frac{R_c}{a}\right)^5.
\label{eq:tidqadim}
\end{eqnarray}
The apsidal motion constant embodies the reaction of the internal structure to the perturbing centrifugal and tidal fields. It is defined as \citep{kopal_dynamics_1978}
\begin{equation}
	k_2 = \frac{3 - \eta_2(R_c)}{2(2+ \eta_2(R_c))},
\end{equation}
where the function $\eta_2(r) = \frac{r}{f_2(r)} \deriv{f_2}{r}$ is solution of Radau's equation and $f_2(r)$ is the surficial distortion by which the perturbed equipotential surfaces of the star can be described with the parametric equation
\begin{equation}
	r'(r, \theta) = r\left(1 + f_0 + f_2(r)P_2(\cos\theta)\right),
\end{equation}
where $r'$ is the perturbed radius, $r$ the unperturbed radius, $f_0$ a constant, $P_2(x) = 3/2x^2 - 1/2$ is the second Legendre  polynomial, and $\theta$ the angle from the axis of the perturbation (i.e. the spin axis and line connecting the two orbiting bodies, for the centrifugal and the tidal deformations, respectively). Thus $f_2$ is a solution of Clairaut's equation. Note that some authors  use the tidal Love number defined as $2k_2$ \citep[e.g.][]{kramm_degeneracy_2011, ogilvie_tidal_2014}.

Expressing the above equations \eqref{eq:rotqadim}-\eqref{eq:tidqadim} as a function of the mass ratio $q$ and the dimensionless Roche-lobe radius $\rho_f$ (equation \eqref{eq:rhof}), the filling factor $f$, and using Kepler's third law $a^3n^2 = G(m_p+m_c)$, we get,
\begin{eqnarray}
\label{eq:rotqadim2}
\rotqadim & = & - \frac{1}{3}k_2\rho_f^5 f^5(1 + q^{-1})  \\
\tidqadim & = & -  k_2 \rho_f^5 f^5  q^{-1} 
\label{eq:tidqadim2}
\end{eqnarray}
Therefore, the relative importance of the quadrupole potential depends only on the mass ratio $q$, the filling factor $f$, and apsidal motion constant $k_2$. Noting that $q \ll 1$ in spiders, we see that $\tidqadim \simeq 3 \rotqadim$. 

The value of $k_2$ is one of the major unknowns of the problem. We note that due to their low mass and high level of irradiation black widows are somewhat similar to so-called hot Jupiters \citep{kramm_constraining_2012} for which large apsidal motion constants have been calculated -- up to $k_2 \sim 0.2$ \citep{kramm_degeneracy_2011, kramm_constraining_2012} -- while sun-like stars have much lower $k_2 \sim 0.015$ (e.g. \citet{ogilvie_tidal_2014}). The latter value is comparable to what is expected in low-mass white dwarfs where $0.01 \lesssim k_2 \lesssim 0.1$ \citep{valsecchi_tidally_2012}. A large fraction of binary pulsars possess white dwarf companions and some redback companions are possibly going to turn into them \citep{bellm_properties_2016}.
On the other hand, spiders might be similar to companions in cataclysmic-variable binaries, where $k_2$ as low as $\sim 10^{-4}$ and $\sim 10^{-3}$  have been estimated in UX UMa and RW Tri respectively \citep{warner_apsidal_1978}. Such low values have been shown to be compatible with the companion being the core of a red giant stripped of its envelope by mass transfer \citep{cisneros-parra_apsidal_1970}. Such mechanism  could well apply to spider systems. It is therefore impossible to estimate precisely in what range the apsidal motion constants of redback and black-widow companions lies, and we can only conjecture $10^{-4} \lesssim k_2 \lesssim 0.2$.

With these limitations in mind we can give an estimate of the relative importance of the quadrupole potential using equations \eqref{eq:rotqadim2}-\eqref{eq:tidqadim2}, 
\begin{equation}
\label{eq:J2final}
	\tidqadim \simeq 3\rotqadim \simeq -4\dix{-4}\left(\frac{k_2}{0.1} \right) \left(\frac{q}{\frac{1}{16}}\right)^{-1} f^5 \left(\frac{\rho_f(q)}{\rho_f(\frac{1}{16})}\right)^{5}.
\end{equation}

\section{Orbital motion with a time-varying quadrupole moment}
 \label{sec:motion}
\newcommand{\rp}{\vec{r}_p}
\newcommand{\rc}{\vec{r}_c}
\newcommand{\vr}{\vec{r}}
\newcommand{\vx}{\vec{x}}
\newcommand{\vy}{\vec{y}}
\newcommand{\vz}{\vec{z}}

\newcommand{\rpdot}{\dot{\vec{r}}_p}
\newcommand{\rcdot}{\dot{\vec{r}}_c}
\newcommand{\vrdot}{\dot{\vec{r}}}

\newcommand{\mr}{m_r}

In addition to the quadrupole field of the companion one must consider that spider systems are relativistic binaries. The relative importance of the first-order relativistic corrections (also called 1PN, for post-Newtonian) to the binary potential can be estimated with
\begin{equation}
\label{eq:relcorrection}
\epsilon = \frac{GM}{ac^2} \sim \frac{v^2}{c^2} \simeq  2\dix{-6} \left(\frac{M}{1.7\Msol}\right)^{2/3} \left(\frac{P}{4\mathrm{h}}\right)^{-2/3},
\end{equation}
where $c$ is the speed of light in vacuum and  $v\simeq 2\pi a/P$ is the characteristic orbital velocity. This implies that relativistic corrections are smaller or comparable to the quadrupole correction given by equation \eqref{eq:J2final}. As we shall see in the next section any higher order relativistic effects are largely undetectable.

It follows that the orbital motion is entirely described by the Hamiltonian
\begin{equation}
H = \frac{1}{2}m_p \rpdot^2 + \frac{1}{2}m_c \rcdot^2 + m_p\Phi_c(\rp - \rc) + \Phi_{1\mathrm{PN}}(\rp - \rc),
\label{eq:hamiltonian}
\end{equation}
where $\Phi_c$ is the potential given by equation \eqref{eq:phic} and $\Phi_{1\mathrm{PN}}$ is the first-order relativistic correction (e.g. \citet{damour_general_1985}) given by general relativity.

Setting the (Newtonian) centre of mass of the system to  $m_p \rp + m_c\rc = \vec{0}$, and using this relation to transform the kinetic terms in equation \eqref{eq:hamiltonian}, one is left with an equivalent one-body problem for $\vr = \rp - \rc$ with reduced mass $\mr = m_pm_c/M$, where $M=m_p +m_c$ is the total mass of the system. The Hamiltonian of the reduced system is
\begin{equation}
H = \frac{1}{2}\mr \vrdot^2  + m_p\Phi_c(\vec{r}) + \Phi_{1\mathrm{PN}}(\vec{r}).
\label{eq:hamiltonianfinal}
\end{equation}
The above Hamiltonian contains the Keplerian two-body motion, the perturbing quadrupole potential proportional to the quadrupole function $\qadim(r,t)$, equation \eqref{eq:qdadimcomponents}, in $\Phi_c$, and the perturbing 1PN potential $\Phi_{1\mathrm{PN}}$.

 At linear order, perturbations can be worked out independently and summed together. In the following we therefore split the problem into two. First, we solve the constant perturbation problem in Section \ref{sec:constantqrelat}, corresponding to the Hamiltonian
\begin{equation}
\label{eq:bH}
\bar{H} = \frac{1}{2}\mr \vrdot^2   - \frac{GM\mr}{r}\left( 1 + \rotqadim\frac{ a^2}{r^2} + \tidqadim\frac{ a^5}{r^5}\right) + \Phi_{1\mathrm{PN}}(\vec{r}).
\end{equation}
Second, we solve the variable-perturbation problem in Section \ref{sec:varproblem}, corresponding to the Hamiltonian
\begin{equation}
\label{eq:vH}
\tilde{H} = \frac{1}{2}\mr \vrdot^2   - \frac{GM\mr}{r}\left( 1 +  \vqadim\frac{ a^2}{r^2}\right).
\end{equation}
In both Hamiltonians, $a$ represents the orbital separation corresponding to the solution of the unperturbed problem. Note that $H \neq \bar{H} + \tilde{H}$, as each Hamiltonian also contains the unperturbed Keplerian two-body motion, but different perturbations.

An additional point to consider is that spiders are characterized by circular or quasi-circular orbits to the point where the eccentricity $e \sim \qadim$. It is therefore not necessary -- and arguably hazardous -- to consider orders in eccentricity larger than the quadrupole or the relativistic correction. Throughout this paper we shall consider the orbital dynamics only to first order in eccentricity $e$, post-Newtonian corrections $\epsilon$, and quadrupole corrections $\qadim$. Cross terms between these small quantities are included only for long-term cumulative effects; i.e. precession (see Section \ref{sec:constantqrelat}).

\subsection{Spin and tidal quadrupole with relativistic effects \label{sec:constantqrelat}} 
As pointed out by \citet{wex_timing_1998}, the non-relativistic problem of a spin quadrupole moment with potential of the form  $-G m_p m_c q/2r^3$, where $q \equiv 2\rotqadim a^2$ is a constant representing the quadrupole moment in Wex's notations (as to being the mass ratio in this paper), can be solved using the same approach that led \citet{damour_general_1985} to a closed-form solution of the relativistic 1PN two-body problem. We will refer to the \citet{damour_general_1985} form as DD solution. In Section \ref{sec:DD} we show how the principle of superposition can be used to incorporate both spin quadrupole effects \citep{wex_timing_1998} and 1PN effects \citet{damour_general_1985} in a single DD solution.

The tidal quadrupole, on the other hand, cannot be solved using the DD approach as the $1/r^6$ dependence of the tidally induced quadrupole potential prevents the use of DD's conchoidal coordinate transformation \citep{damour_general_1985}. However, it can be solved to first order using the method of variation of the elements (see appendix \ref{secap:pseudoDD} and Section \ref{sec:varproblem} below), leading to a solution in a quite unwieldy form. We show in appendix \ref{secap:pseudoDD} that, to first order in eccentricity, the solution obtained through the latter method can be matched with the form of a DD solution through a redefinition of the orbital elements. We will refer to such a solution as an effective-DD solution.

We conclude that the entire problem of a synchronised and quasi-circularised orbit, including the effects of spin and tidal deformations as well as 1PN relativistic corrections, can be cast into the simple form of an effective-DD solution. 

Such a solution is described by the orbital elements (e.g \citet{beutler_methods_2004}): \\
$a$, the semi-major axis, \\
$e$, the eccentricity,\\
$i$, the inclination of the orbital plane, \\
$\Omega$, the longitude of ascending node, \\
$\omega$, the longitude of periastron, \\
$T_p$, the time of passage at periastron,\\
as well as by what we shall call post-Keplerian elements $k,\delta,\delta_r,\delta_v$ in reference to, but different from, post-Keplerian parameters \citep{damour_strong-field_1992}, and which we describe in the following subsection.

These DD orbital elements are a convenient way of representing the mathematical solution of the orbit in a quasi-Keplerian way. However, they are not equal to the osculating orbital elements describing the unperturbed Keplerian orbit that is instantaneously tangent the actual trajectory. We provide the relation between the two sets of parameters in appendix \ref{secap:osculating}. 

\subsubsection{The DD and effective-DD solution \label{sec:DD}}
\begin{figure}
	\includegraphics[width=0.5\textwidth]{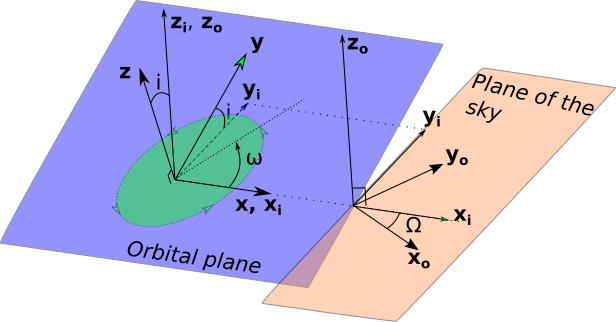}
	\caption{Representation of the relations between the frame of the observer $(\vx_o, \vy_o,\vz_o)$, the intermediate frame $(\vx_i, \vy_i,\vz_i)$ and the frame of the orbital plane $(\vx, \vy,\vz)$, formally given by equations \eqref{eq:rotfirst}-\eqref{eq:rotlast}. The observer is looking in the direction $\vz_o$, while $\vz$ is perpendicular to the orbital plane.\label{fig:frame}}
\end{figure}
As pointed out above, the DD solution is not restricted to relativistic effects but can be applied to a whole class of perturbation problems. We use this approach here to perturbatively solve the problem described in equation \eqref{eq:bH}.

 As in the Keplerian problem, the perturbed motion is planar and we work here in the frame given by the direct triad $(\vx, \vy, \vz)$ with origin at the centre-of-mass of the system (as defined in \citet{damour_general_1985}), where $\vx$ points towards the ascending node, and $\vz$ in the direction of the orbital angular momentum. In general, the motion can be expressed in the frame of the observer $(\vx_o, \vy_o, \vz_o)$, where $\vz_o$ gives the direction from the observer to the centre-of-mass. As shown in Figure \ref{fig:frame}, one performs a rotation of angle $i$ to the intermediate frame $(\vx_i, \vy_i, \vz_i)$ followed by a rotation of angle $\Omega$,
\begin{eqnarray}
\label{eq:rotfirst}
	\vy & = & \cos i \vy_i + \sin i \vz_i, \\
	\vz & = &  -  \sin i \vy_i + \cos i \vz_i,
\end{eqnarray}
where $\vx = \vx_i$, and 
\begin{eqnarray}
\vx_i & = & \cos \Omega \vx_o + \sin \Omega \vy_o, \\
\vy_i & = & -  \sin \Omega \vx_o + \cos \Omega \vy_o ,
\label{eq:rotlast}
\end{eqnarray}
where $\vz_i = \vz_o$.

The DD solution to the reduced one-body problem \eqref{eq:bH} is formally very similar to the Keplerian two-body motion. It is entirely contained in the following set of equations:
\begin{eqnarray}
	\label{eq:DDpos}
	\vec{r} & = & r(\cos v, \sin v,)_{(\vx, \vy, \vz)} \\
	\label{eq:DDdist}
	r & = & a(1-e_r\cos E), \\
	\label{eq:DDv}
	v & = & \omega + (1+k) 2\arctan\left(\sqrt{\frac{1+e_v}{1-e_v}}\tan \frac{E}{2}\right), \\
	\label{eq:DDkepler}
	E - e \sin E & = & n (t - T_p), \\
	a^3 n^2 & = &  GM\left(1+\delta\right).
	\label{eq:DD3rdlaw}
\end{eqnarray}
The effect of the perturbation is to introduce the eccentricity parameters $e_v = e(1+\delta_v), e_r = e(1+\delta_r)$, the precession constant $k$, and modify Kepler's third law with $\delta$ such that the post-Keplerian elements $k\sim\delta\sim\delta_r\sim\delta_v$ scale with the perturbation. For simplicity, in the following, we will write that perturbations are generally of order $\sim k$.

In the case where $e_r = e_v = e$ and $k=\delta=0$ one recognises the equations of the Keplerian reduced problem with total mass $M$, semi-major axis $a$, eccentricity $e$, angular position $v$, eccentric anomaly $E$, mean motion $n=2\pi/P$ and time of periastron passage $T_p$. Equation \eqref{eq:DDpos} gives the position of the body with respect to the centre-of-mass of the system, equation \eqref{eq:DDkepler} is Kepler's modified equation, and equation \eqref{eq:DD3rdlaw} is Kepler's modified third law.  

The post-Keplerian elements for different perturbations of similar order of magnitude can be added linearly. Besides, as we further restrict our study to low-eccentricity systems the difference between the three eccentricity parameters can be neglected since it is of order $\sim ek$. It follows that, in this work,
\begin{eqnarray}
\label{eq:k}
	k & = & 3\rotqadim  + (15\tidqadim)_{\mathrm{eff}} + 3\epsilon, \\
\label{eq:delta}
	\delta & = & - 3\rotqadim - (15\tidqadim)_{\mathrm{eff}} + \frac{\epsilon}{2}\left(\frac{m_cm_p}{M} - 9\right) +\bigcirc(e^2) ,
\end{eqnarray}
where the $\epsilon$ terms give the 1PN corrections \citep{damour_general_1985}, the $\rotqadim$ terms give the  spin quadrupole correction \citep{wex_timing_1998} and the $\tidqadim$ terms are the effective contributions that allow the inclusion of the tidal quadrupole within the DD formalism to first order in eccentricity (see appendix \ref{secap:pseudoDD}). The ``eff'' subscript is here to remind us that it is obtained through the matching procedure of appendix \ref{secap:pseudoDD}, using equation \eqref{apeq:pseudoDDkdelta}. Besides, to leading order in $ek$,
\begin{equation}
		e_v = e_r =  e.
\end{equation}

Of particular interest for timing is the orbit of the pulsar itself. Using the centre-of-mass relation of the perturbed problem \citep{damour_general_1985} one finds that the angular motion is unchanged compared to the reduced problem of equations \eqref{eq:DDpos}-\eqref{eq:DD3rdlaw}, and the radial motion is given by 
\begin{equation}
	r_p = a_p \left(1 - e_{r_p}\cos E\right),
\end{equation}
where the pulsar semi-major axis $a_p = a \frac{m_c}{M}$, and as before $e_{r_p} = e + \bigcirc(ke)$. Therefore, at leading order in $ke$ the centre-of-mass relation of a DD solution is identical to the Keplerian one.

\subsection{Variable quadrupole $\vqadim$}
\label{sec:varproblem}
The variable quadrupole component $\vqadim$ generates perturbations which, as we will see below, are finite to first order. Therefore, the variable problem \eqref{eq:vH} can be readily treated to first-order using the method of variation of the elements.

\subsubsection{Variation of the elements}
\newcommand{\Phip}{\Phi_p}
The perturbation of the two-body Keplerian motion by a perturbing potential $\Phip$ results in variations of the orbital elements with time. The orbital elements are defined here as the elements describing the osculating orbit, namely the tangential Keplerian orbit, at the time of passage at ascending node $T_a$. We keep here the same notations as in previous section, though the orbital elements $a, e, i,\Omega,\omega,T_p$ differ at first order from their DD counterpart (see appendix \ref{secap:osculating}). We will match the two approaches in Section \ref{sec:match}.

The time evolution is described to first order by the so-called Lagrange planetary equations \citep[e.g.][]{beutler_methods_2004,wex_timing_1998}:
\begin{eqnarray}
\label{eq:llp}
	\dot{p} & = & \frac{2\sqrt{1- e^2}}{na} \pderiv{\Phip}{\omega}, \\
	\dot{e} & = & -\frac{\sqrt{1-e^2}}{na^2 e}\left[ \pderiv{\Phip}{\omega} +\frac{\sqrt{1- e^2}}{n}\pderiv{\Phip}{T_0} \right], \\
	\dot{i} & = & \frac{1}{na^2 \sqrt{1-e^2}} \left[ \frac{1}{\tan i} \pderiv{\Phip}{\omega} - \frac{1}{\sin i}\pderiv{\Phip}{\Omega} \right], \\
	\dot{\Omega} & = & \frac{1}{na^2 \sqrt{1 - e^2} \sin i} \pderiv{\Phip}{i}, \\
	\dot{\omega} & = & \frac{\sqrt{1 - e^2}}{na} \left[ -2 \pderiv{\Phip}{p} + \frac{1}{ae} \pderiv{\Phip}{e} - \frac{1}{p \tan i} \pderiv{\Phip}{i} \right], \\ 
	\dot{T}_p & = & \frac{1 - e^2}{n^2 a^2 e} \pderiv{\Phip}{e},
\label{eq:llT0}
\end{eqnarray}
where $p=a(1-e^2)$ is the semi-latus rectum, $n = 2\pi/P = \sqrt{\mr/a^3}$ is the mean motion, and the perturbing potential is given by the $\vqadim$ term of \eqref{eq:vH},
\begin{equation}
	\Phip = \vqadim(t) \frac{GM}{r}\frac{a^2}{r^2},
\end{equation}
such that the perturbing acceleration is $\vec{F_p}=\mr\vec{\nabla}\Phip$ (note the absence of minus sign in the present convention).

\newcommand{\ks}{\kappa_s}
\newcommand{\kc}{\kappa_c}
Equations \eqref{eq:llp}-\eqref{eq:llT0} are somewhat impractical for low eccentricities, particularly due to the $1/e$ factors in some terms. We therefore adopt a different parametrisation in the rest of this paper, defining the so-called Laplace-Lagrange parameters \citep{lange_precision_2001}.
\begin{eqnarray}
\label{eq:kc}
	\kc & = & e\cos\omega \\
	\ks & = & e\sin\omega
\label{eq:ks}
\end{eqnarray}
instead of $e$ and $\omega$, and using the time of passage at ascending node,
\begin{equation}
\label{eq:Ta}
T_a = T_p - \frac{\omega}{n} + 2\frac{\ks}{n} + \bigcirc\left(e^2\right),
\end{equation}
instead of $T_p$ as fiducial time reference. Equations for $\dot{\kc}, \dot{\ks}, \dot{T_a}$ are obtained after taking the time derivative of equations \eqref{eq:kc}-\eqref{eq:Ta} and inserting equations \eqref{eq:llp}-\eqref{eq:llT0}.

\subsubsection{Results: the perturbed orbital elements}
The perturbed orbital elements are
\begin{equation}
\label{eq:elementperturbed}
x^{(1)}(t) = x + \Delta x(t),
\end{equation}
where $x\in\{a,\kc,\ks,i,\Omega,\omega,T_a\}$ an unperturbed element and $\Delta x(t) = \int_{T_a}^{t}\dif{t'} \dot{x}(t')$.
The instantaneous value of the perturbed orbital elements $x^{(1)}(t)$ defines the so-called osculating orbit to the binary motion, e.g. the unperturbed Keplerian orbit instantaneously tangential to the actual orbit.

We obtain  $\Delta i = \Delta \Omega = \Delta a = 0$ and
\begin{eqnarray}
\label{eq:dkc}
\Delta\kc & = & -3 n \Delta_s, \\
\label{eq:dks}
\Delta\ks & = & 3 n \Delta_c, \\
\Delta T_a & = & 6 \left(\Delta_c - \Delta_1 \right), 
\label{eq:dta}
\end{eqnarray}
where we have introduced 
\begin{eqnarray}
\label{eq:Deltac}
	 \Delta_c & = & \int_{T_a}^{t}\dif{t'} \vqadim(t') \cos\sigma', \\
	 \label{eq:Deltas}
	 \Delta_s & = & \int_{T_a}^{t}\dif{t'} \vqadim(t')\sin\sigma', \\
	 \Delta_1 & = & \int_{T_a}^{t}\dif{t'} \vqadim(t'),
\end{eqnarray}
with $\sigma' = n(t'-T_a)$.

\newcommand{\vDelta}{\tilde{\Delta}}
We shall recall that the time scale for quadrupole variations is $T_2>> P$, and therefore $\Delta_c \sim \Delta_s \sim P$ and $\Delta_1 \sim T_2$. As a result $\Delta_1/\Delta_c \sim T_2/P >> 1$ in $\Delta T_a$ \eqref{eq:dta}, while the variations of eccentricity are of order $\sim 3\vqadim(t)$ in $\Delta \kc, \Delta\ks$ (equations \eqref{eq:dkc}-\eqref{eq:dks}).

It is worth mentioning that we could readily calculate the perturbation terms at order $e\vqadim$, had we not decided to limit ourselves to lower order eccentricity terms. We have checked that no secular effect would become important at order $e\vqadim$ and found that, despite the fact that terms linearly growing with time are present, their effect identically cancels once injected in a Keplerian model and thus need not be included here.

\subsection{Matching with the constant problem}
\label{sec:match}
In Section \ref{sec:varproblem}, we calculated the orbital elements corresponding to the osculating orbit at the time of passage at ascending node $T_a$, while in Section \ref{sec:constantqrelat} we defined them through the DD formalism. One can show that, as expected, the two sets of orbital elements are identical to zeroth order in perturbation (see appendix \ref{secap:osculating}). It follows that the DD elements can be used interchangeably in equations \eqref{eq:llp}-\eqref{eq:llT0}, which remain accurate to first order. 

Consequently, the solution to the general problem \eqref{eq:hamiltonianfinal} is obtained by merely replacing the DD orbital elements $x\in\{a,\kc,\ks,i,\Omega,\omega,T_a\}$ by $x \rightarrow x + \Delta x$, equation \eqref{eq:elementperturbed}, using equations \eqref{eq:dkc}-\eqref{eq:dks}.

\section{Timing model for spider pulsars }
\label{sec:timing}

\subsection{Orders of magnitude of spider timing}
In this subsection we will briefly review the fundamental principles behind binary pulsar timing in order to establish the order of magnitude of its various components in spider systems. Pulsars have a rotational phase $\phi(\tau)$ that evolves very regularly with the proper time of the pulsar $\tau$, to the point that the phase can usually be expanded merely to second order in time,
\begin{equation}
\label{eq:psrturn}
	N(\tau) = \nu\tau + \frac{1}{2}\dot{\nu}\tau^2,
\end{equation}
where $N=\phi/2\pi$ is the turn number, $\nu$ the spin frequency, and $\dot{\nu}$ is the frequency derivative. In pulsars, the derivative is related to the pulsar spin-down due to the conversion of rotational energy into pulsar wind and electromagnetic emissions in the magnetosphere. In a few cases (see \citet{marshall_new_2016} and references therein) a second derivative could be measured. However, spider pulsars are recycled millisecond pulsars \citep{roberts_surrounded_2012} with a characteristically low spin-down, making them even more unlikely to have a detectable second derivative. 

On its path to telescopes, the signal suffers a number of delays which, when they are time or frequency dependent break the regularity and simplicity of equation \eqref{eq:psrturn}. It is customary to use the formal time of arrival at the Solar system barycentre (SSB) at infinite frequency as an intermediate between delays due to the Solar system and delays due to the pulsar binary system and propagation through interstellar space \citep{blandford_arrival-time_1976, damour_general_1986, wex_timing_1998, voisin_simulation_2017},
\newcommand{\tassb}{t_a^{\mathrm{SSB}}}
\begin{equation}
	\tassb = \tau_e + \Delta_R + \Delta_E + \Delta_S + \Delta_K + \Delta_A + \Delta_{\mathrm{DM}} + \Delta_D.
\end{equation}
where $\tau_e$ is the time of emission in the pulsar frame.

The Doppler delay $\Delta_D$ is purely an effect of proper motion and acceleration of the binary with respect to the Solar system, and is usually neglected because unseparable from $\nu$ and $\dot{\nu}$. It results from the variation of the Doppler factor between the SSB and the binary centre-of-mass frame. It is totally independent of binary motion, assuming general relativity to be correct, and therefore we refer the interested reader to the literature (e.g. \citet{edwards_tempo2_2006, damour_strong-field_1992}).
 
The dispersion measure (DM) delay $\Delta_{\mathrm{DM}}$ results from the scattering of radio waves on free electrons of the interstellar medium. Once this delay removed, the time of arrival is virtually at  \textquotedblleft infinite frequency\textquotedblright. In spider pulsars, huge dispersion measure variations are seen (e.g. \citet{polzin_low-frequency_2018}) particularly around superior conjunction. Therefore, unlike in other pulsars this delay is strongly modulated by orbital motion. It can in principle be fitted out by using data in several frequency bands, and/or by trimming the data around the eclipse. Alternatively, gamma rays do not interact significantly with the interstellar medium and therefore gamma-ray data is not sensitive to this effect. We will not consider it further. 

The aberration delay $\Delta_A$ is the relativistic aberration of the pulsar beam due to the velocity of the frame of the pulsar relative to the observer. Its amplitude is $\sim\frac{a_p}{\nu Pc} \sim 2 \frac{m_c}{0.1\Msol} \left(\frac{M}{1.7\Msol}\right)^{-2/3} \left(\frac{P}{4}\right)^{-1/3} \frac{250\mathrm{Hz}}{\nu} \mathrm{ns}$ where $a_p = a m_c/M$ is the pulsar semi-major axis. This is not detectable with current facilities and shall be dismissed below.  

The Kopeikin delays $\Delta_K$ account for the parallax effect arising from the motion of amplitude $\leq a_p$ of the pulsar on the plane of the sky at a finite distance $d$, which can couple with the Earth's own motion of amplitude $a_E \simeq 1$AU \citep{kopeikin_possible_1995, kopeikin_proper_1996}. Given the close orbit of spiders $a_p << a_E$ and the Kopeikin delay is $ \sim \frac{a_p a_E}{cd} \sim 10 \frac{m_c}{0.1\Msol} \left(\frac{M}{1.6\Msol}\right)^{-2/3} \left(\frac{P}{4}\right)^{2/3} \frac{100\mathrm{pc}}{d} \mathrm{ns}$. One last term accounts for the change of viewing geometry due to the proper motion at velocity $V$ of the binary on the plane of the sky relative to the Solar system. This delay grows linearly with time at a rate $a_p V /cd \sim 1 \frac{V}{500\mathrm{km/s}}\frac{m_c}{0.1\Msol} \left(\frac{M}{1.7\Msol}\right)^{-2/3} \left(\frac{P}{4}\right)^{2/3} \frac{100\mathrm{pc}}{d} \mathrm{\mu s}/\mathrm{yr}$. Therefore the Kopeikin terms are negligible unless the pulsar has a particularly fast proper motion and is exceptionally nearby. In the following these terms will therefore be dismissed. 

The Shapiro delay $\Delta_S$ accounts for the relativistic distortion of the light path due to the gravitational field of the companion. It is typically of order $Gm_c/c^3 \sim 0.5 \frac{m_c}{0.1\Msol}  \mathrm{\mu s}$ except in the unlikely event that the orbital inclination is extremely close to $90^\circ$, but in this case the beam would be intercepted by the companion. Therefore, even in the unlikely event that this delay can be detected, its calculation based on the unperturbed two-body dynamics would be accurate enough, and we will no longer consider it in this paper. 

The Einstein delay $\Delta_E$ connects the coordinate time at infinity to the proper time of the pulsar, by including the distortion due to the gravitational field of the companion and the velocity of the pulsar. Its time-variable component (the only one detectable), is of order $eP \frac{Gm_c}{2\pi ac^2} \sim 3\frac{e}{10^{-5}}\frac{m_c}{0.1\Msol} \left(\frac{P}{4\mathrm{h}} \frac{M}{1.7\Msol}\right)^{1/3}  \mathrm{n s}$. Likewise the Shapiro delay, this delay is unlikely to be detectable in current datasets, and even if it were a calculation based on unperturbed motion would be accurate enough, so we shall dismiss it in the following. 

The R\oe mer delay $\Delta_R$ is the most important delay for pulsars in binary systems. It accounts for the geometric variation of the light travel time across the orbit. It is of order $ \sim a_p\sin i/c \sim 0.2 \sin i\frac{m_c}{0.1\Msol} \left(\frac{M}{1.7\Msol}\right)^{-2/3} \left(\frac{P}{4\mathrm{h}}\right)^{2/3} \mathrm{s}$. As we shall see below, this is the only delay significantly affected by quadrupole effects.

\subsection{The perturbed R\oe mer delay}

The Roemer delay is the time taken by light to cross the distance between the pulsar and the binary barycentre projected onto the line connecting the latter to the Solar system barycentre,
\newcommand{\kssb}{\vz_o}
\begin{equation}
\label{eq:roemerbase}
	\Delta_R = \frac{\kssb\cdot \vr_p}{c},
\end{equation}
where $\kssb$ is a unit vector along the line going from the Solar system barycentre to that of the binary (equations \eqref{eq:rotfirst}-\eqref{eq:rotlast}), and $\vr_p$ is the pulsar position relative to the binary barycentre.

\subsubsection{Low eccentricity, constant quadrupole and relativistic effects}
For low eccentricity binaries, it has been pointed out \citep{lange_precision_2001} that a parametrisation in terms of the Laplace-Lagrange parameters $\kc, \ks$, equations \eqref{eq:kc}-\eqref{eq:ks}, to first order in eccentricity is more adapted than the parametrisation in terms of eccentricity and longitude of periastron $e,\omega$. This is essentially because the longitude of periastron is poorly constrained in a quasi-circular orbit, which mathematically leads to degeneracies between the parameters. Thus, we propose here a generalisation of the low-eccentricity binary R\oe mer delay proposed in \citep{lange_precision_2001} and implemented as the ELL1 model in the timing software Tempo2 \citep{hobbs_tempo2_2006,edwards_tempo2_2006}, which is based on a DD solution of the equations of motion rather than a purely Keplerian solution. 

Combining equations \eqref{eq:DDkepler} and \eqref{eq:DDv}, one obtains the time of passage at periastron to first order in eccentricity, 
\begin{equation}
T_a = \bar{T}_a + \frac{(e+e_v)}{\bar{n}} \sin\omega + \bigcirc(e^2),
\end{equation}  
where $\omega$ is the longitude of periastron, 
\begin{equation}
\label{eq:obsmeanmotion}
\bar{n}=n(1+k)
\end{equation} 
is the observable mean motion, and
\begin{equation}
\label{eq:taobs}
\bar{T}_a = T_p - \frac{\omega}{\bar{n}}
\end{equation}
is the observable time of passage at ascending node.

\newcommand{\sigmab}{\sigma}

Using the fact that $E = n(t-T_p) + e\sin n(t-T_p) + \bigcirc(e^2)$ (equation \eqref{eq:DDkepler}), combined with equation \eqref{eq:taobs},
it follows that
\begin{equation}
\label{eq:Ee2}
E = \sigmab -\omega + e\sin(\sigmab - \omega) + \bigcirc(e^2)
\end{equation}
where $\sigmab = \bar{n}\left(t- \bar{T}_a\right)$.

Inserting \eqref{eq:Ee2} in \eqref{eq:DDv} one also obtains the angular position,
\begin{equation}
\label{eq:ve2}
v = \sigmab  + (e + e_v) \sin \left((1-k)(\sigmab - \omega)\right) + \bigcirc(e^2).
\end{equation} 

Inserting \eqref{eq:Ee2}-\eqref{eq:ve2} in \eqref{eq:DDpos}-\eqref{eq:DDdist}, expanding consistently to first order in eccentricity and in perturbation with the restriction that secularly growing terms of order $e k\sigma$ must be kept, and inserting the result in equation \eqref{eq:roemerbase}, one obtains the R\oe mer delay,
\begin{equation}
\label{eq:roemer}
\Delta_R = x\left( \sin\sigmab -\frac{3}{2} \ks' - \frac{\ks'}{2}\cos 2\sigmab + \frac{\kc'}{2}\sin 2\sigmab  \right)
\end{equation}
where $x = a_p\sin i /c$ is the projected semi-major axis of the pulsar orbit, and 
\begin{eqnarray}
\label{eq:kcp}
\kc' & = & e\cos\left(\omega + k \sigmab\right), \\
\label{eq:ksp}
\ks' & = & e\sin\left(\omega + k \sigmab\right) 
\end{eqnarray}
are the precessing Laplace-Lagrange parameters which can be readily expressed as a function of the Laplace-Lagrange parameters $\kc, \ks$ (equations \eqref{eq:kc}-\eqref{eq:ks}) by expanding the trigonometric functions.

Thus, we find that the generalised ELL1 model is obtained by replacing the Laplace-Lagrange parameters of \citet{lange_precision_2001} by their precessing counterparts \eqref{eq:kcp}-\eqref{eq:ksp}. This results in modulating the eccentricity terms with an envelope at the precession angular frequency
\begin{equation}
\label{eq:omdot}
	\dot{\omega} = k\bar{n}.
\end{equation}
Additionally, we note that the second term of \eqref{eq:roemer} had no counterpart in the original ELL1 model: although it is mathematically present, it had been discarded as an unconstraining constant. Here, this terms oscillates with the precession angular frequency $\dot{\omega}$. Such slow evolution is prone to be absorbed in timing-noise removal \citep[see, e.g.,][]{coles_pulsar_2011,van_haasteren_understanding_2013}, in particular if the removal is not done simultaneously and self-consistently with the fit of the timing model. Thus, if timing noise removal is not performed self-consistently, it could be more accurate to remove the second term from \eqref{eq:roemer}, being understood that it is fitted as a component of timing noise. 

Equation \eqref{eq:roemer} was also found in \cite{susobhanan_exploring_2018} although these authors did only consider secular precession, and not short-term  periodic effects. Here, these effects are effectively taken into account by the redefinition of the orbital parameters within the DD framework (see also appendix \ref{secap:osculating}). We shall see in Section \ref{sec:eccperiastronprec} that this has the very important consequence that the minimum eccentricity of the orbit is not zero but a value strictly larger. Further, we account for the effects of a variable quadrupole moment.

\subsubsection{Complete perturbed R\oe mer delay with variable quadrupole}
\label{sec:completeroemer}

We can now operate the replacement $X \rightarrow X + \Delta X$ where $X \in \{T_a,\kc,\ks\}$ using equations \eqref{eq:dkc}-\eqref{eq:dta} in order to obtain the perturbed R\oe mer delay from equation \eqref{eq:roemer}. We should keep in mind that this procedure is only valid at linear order in the perturbation parameters.  

Noting that the time scale of quadrupole variations is much longer than the orbital period, $T_2 \gg P$, we evaluate equations \eqref{eq:Deltac}-\eqref{eq:Deltas} as
\begin{eqnarray}
n\Delta_c & = & \vqadim(t)\sin\sigmab + \bigcirc(P/T_2), \\
n\Delta_s & = & -\vqadim(t)\cos\sigmab + \bigcirc(P/T_2).
\end{eqnarray}

With that simplification, the variable quadrupole perturbation is effectively obtained from equation \eqref{eq:roemer} by performing the replacements
\begin{eqnarray}
	\sigma & \rightarrow & \sigma'  =  \sigma + 6\bar{n}\int_{T_a}^{t}\dif{t'}\vqadim(t'), \\
	x & \rightarrow & x' = x - 3\vqadim(t).
	\label{eq:xprime}
\end{eqnarray}
such that our final R\oe mer delay is given by 
\begin{equation}
\label{eq:finalroemer}
\Delta_R = x'\left( \sin\sigma' -\frac{3}{2} \ks' - \frac{\ks'}{2}\cos 2\sigma' + \frac{\kc'}{2}\sin 2\sigma'  \right),
\end{equation}
with the understanding that for terms at first order in eccentricity $\sigma$ and $\sigma'$ are equivalent.

The main effect of the variable quadrupole lies in the replacement of the orbital phase $\sigma \rightarrow \sigma'$. Indeed, Taylor expanding equation \eqref{eq:finalroemer} to first order in $\vqadim$, we find that $\Delta_R = \Delta_R(x,\sigma) + \Delta_R^\sigma + \Delta_R^x$ where $\Delta_R(x,\sigma)$ is given by equation \eqref{eq:roemer} and 
\begin{eqnarray}
\label{eq:deltaRsigma}
	\Delta_R^\sigma & = & 6x\bar{n}\int_{T_a}^{t}\dif{t'}\vqadim(t') \cos\sigma, \\
	\Delta_R^x &  = & -3x\vqadim(t)\sin\sigma.
\label{eq:deltaRx}
\end{eqnarray}
It follows that $\Delta_R^\sigma / \Delta_R^x \sim 3T_2/P \gg 1$.

Orbital period variability in spiders is usually accounted for by a Taylor expansion \citep[e.g.][]{pletsch_gamma-ray_2015},
\begin{equation}
\label{eq:nderivatives}
\sigma'(t) =  \sum_{i=0} \frac{\bar{n}^{(i)}}{(i+1)!} \left(t - \bar{T}_a \right)^{i+1} 
\end{equation} 
where $\bar{n}^{(0)} = \bar{n}$ is the observable orbital angular frequency, and $n^{(i)}$s are empirical parameters of the timing model.

It follows from this equation that the instantaneous orbital period is $P(t) = P + \Delta P(t)$, with
\begin{equation}
\label{eq:dp}
\frac{\Delta P(t)}{P} = 1 - \frac{1}{\bar{n}}\deriv{\sigma'}{t} = -6\vqadim(t) + \bigcirc(P/T_2),
\end{equation}
where at this order $\bar{n}$ and $n$ are equivalent. We emphasise that this result holds only because $T_2 \gg P$ and corresponds exactly to the results derived by \citet{applegate_mechanism_1992} from conservation laws.

We now turn to the apparent change in projected semi-major axis given by $x \rightarrow x'$. This effect was noted in \citep{lazaridis_evidence_2011}, considering the \citet{applegate_mechanism_1992} mechanism. These authors pointed out that, since the orbital angular momentum is conserved in this model, a change in orbital period leads to a change in separation through Kepler's third law. Our present model also conserves angular momentum, and we therefore obtain the same effect. However we note that \citet{lazaridis_evidence_2011} gave the uncorrect relation $\Delta a/a = 2\Delta P /P$ while we have from \eqref{eq:xprime} and \eqref{eq:dp} $2\Delta a/a = \Delta P /P$. In any case, we see that given the order of magnitude of $\vqadim$ (see next section) this effect is not detectable in known pulsars, as also concluded \citet{lazaridis_evidence_2011}, as it would only contribute $\sim x\vqadim(t)$ to the Roemer delay.

\section{Discussion}
\label{sec:application}
\newcommand{\emin}{e_{\mathrm{min}}}

We report in Table \ref{tab:pulsars} the list of spider pulsars for which orbital period variations have been observed so far, together with estimates of some relevant quantities discussed in this paper. In particular, we discuss corrections to mass estimates in Section \ref{sec:masscorrection}, periastron precession in Section \ref{sec:eccperiastronprec}, and period variations in Section \ref{sec:discussvariations}. 
\newcommand{\omdot}{\dot{\omega}}
\newcommand{\normvqadim}{\norm{\vqadim}}
\begin{table*}
		\caption{Spider pulsars for which orbital period variations have been observed so far along with the main quantities discussed in this paper. Derived from the literature: estimated companion mass $m_c$; amplitude of the quadrupole variations $\normvqadim$, equation \eqref{eq:normqadim}; time scale of period variations $T_2$; estimate of the amplitude of the timing delay associated with the period variation itself $\Delta_R^\sigma \sim 6x\vqadim $, equation \eqref{eq:deltaRsigma}; estimate of the amplitude of the timing delay associated with the variation of the projected semi-major axis $\Delta_R^x \sim 3x\vqadim$, equation \eqref{eq:deltaRx}. 
		Predicted: periastron precession $\dot{\omega}$ assuming the companion is Roche lobe filling, $f=1$, and $k_2=0.1$, equation \eqref{eq:omdot}; minimum eccentricity \eqref{eq:emin} ; amplitude of the delay associated with $\emin$, $\Delta_R^{\emin} \sim x\emin$. Interline: idem assuming $k_2 =0.01$ and $f=0.5$.\label{tab:pulsars} }

\begin{tabular}{ l  l | l l l l | l l l c}
	\multicolumn{2}{c |}{System} & \multicolumn{4}{c|}{Variable quadrupole} & \multicolumn{3}{c}{Precession \& Minimum eccentricity} & References \\
	\hline
   PSR & $m_c$ ($\Msol$) & $\tilde{J}_2 (10^{-8})$ & $T_2$(yr) & $\Delta_R^{\sigma}$ (ms) & $\Delta_R^{x}$ ($\mu$s) & $\dot{\omega}$ (cyc/yr) & $e_{\mathrm{min}}$ & $\Delta_R^{e_{\mathrm{min}}}$ ($\mu$s)  &  \\
   \hline
  J2051-0827  & 0.05 & 5.8 & 6.2 & 0.36 & 0.0079 & -12 & 0.0032 & 1.4e+02 & [1] \\
 &  &  &  &  &  & \ \ \ \ \ \ \ \ 0.0027 & \ \ \ \ \ \ \ \ 9.9e-06 & \ \ \ \ \ \ \ \ 0.45 &  \\
J1959+2048  & 0.03$^a$ & 3.1 & 19$^b$ & 0.29 & 0.0083 & -2.7 & 0.0028 & 2.5e+02 & [2] \\
 &  &  &  &  &  & \ \ \ \ \ \ \ \ -0.005 & \ \ \ \ \ \ \ \ 8.8e-06 & \ \ \ \ \ \ \ \ 0.78 &  \\
J1731-1847  & 0.039$^a$ & 5.7 & 17$^b$ & 0.81 & 0.02 & -3.8 & 0.0032 & 3.9e+02 & [3] \\
 &  &  &  &  &  & \ \ \ \ \ \ \ \ -0.0071 & \ \ \ \ \ \ \ \ 1e-05 & \ \ \ \ \ \ \ \ 1.2 &  \\
J0024-7204J  & 0.03$^a$ & 0.29 & 1.2e+02$^c$ & 0.26 & 0.00035 & -8.5 & 0.0028 & 1.1e+02 & [4] \\
 &  &  &  &  &  & \ \ \ \ \ \ \ \ -0.0033 & \ \ \ \ \ \ \ \ 8.8e-06 & \ \ \ \ \ \ \ \ 0.35 &  \\
J0024-7204O  & 0.01$^a$ & 1.8 & 22$^b$ & 0.29 & 0.0024 & -4 & 0.0015 & 67 & [5] \\
 &  &  &  &  &  & \ \ \ \ \ \ \ \ 0.0065 & \ \ \ \ \ \ \ \ 4.6e-06 & \ \ \ \ \ \ \ \ 0.21 &  \\
J1807-2459A  & 0.01$^a$ & 0.65 & 61$^c$ & 0.15 & 0.00024 & -7.5 & 0.0015 & 18 & [6] \\
 &  &  &  &  &  & \ \ \ \ \ \ \ \ 0.032 & \ \ \ \ \ \ \ \ 4.6e-06 & \ \ \ \ \ \ \ \ 0.056 &  \\
J2339-0533  & 0.32 & 8.4 & 6.2 & -3.6 & 0.16 & -16 & 0.0084 & 5.2e+03 & [7] \\
 &  &  &  &  &  & \ \ \ \ \ \ \ \ -0.037 & \ \ \ \ \ \ \ \ 2.6e-05 & \ \ \ \ \ \ \ \ 16 &  \\
J1023+0038  & 0.2 & 8.5 & 35$^c$ & 11 & 0.088 & -12 & 0.0065 & 2.2e+03 & [8] \\
 &  &  &  &  &  & \ \ \ \ \ \ \ \ -0.025 & \ \ \ \ \ \ \ \ 2e-05 & \ \ \ \ \ \ \ \ 7 &  \\
J1723-2837  & 0.4 & 31 & 2.7$^b$ & 3.7 & 1.1 & -5.6 & 0.0094 & 1.1e+04 & [9] \\
 &  &  &  &  &  & \ \ \ \ \ \ \ \ -0.016 & \ \ \ \ \ \ \ \ 2.9e-05 & \ \ \ \ \ \ \ \ 36 &  \\
	\multicolumn{10}{l}{$^a$Assuming a pulsar mass of $1.35\Msol$ and an inclination of 60\textdegree.}  \\ 
	\multicolumn{10}{l}{$^b$ Observation time span shorter than $T_2$, but at least two orbital frequency derivatives.} \\
	\multicolumn{10}{l}{$^c$ Only one orbital frequency derivative detected, likely due to an observation time span too short: $T_2$ is unreliable.} \\
	\multicolumn{10}{l}{$^d$ An alternate solution is $m_c = 0.7\Msol$ \citep{van_staden_active_2016}. } \\
	\multicolumn{10}{l}{[1]: \citet{lazaridis_evidence_2011, shaifullah_21_2016}; [2]: \citet{arzoumanian_orbital_1994}; [3]: \citet{ng_high_2014}; [4]: \citet{freire_further_2003}; } \\
	\multicolumn{10}{l}{[5]: \citet{lynch_timing_2012}; [6]: \citet{pletsch_gamma-ray_2015}; [7]: \citet{archibald_radio_2009}; } \\
	\multicolumn{10}{l}{[8]: \citet{crawford_psr_2013, van_staden_active_2016}}
\end{tabular}
\end{table*}

\subsection{Mass measurements}
\label{sec:masscorrection}
\newcommand{\bn}{\bar{n}}
As shown in Section \ref{sec:completeroemer}, the orbital period effectively observable through the R\oe mer delay, equation \eqref{eq:finalroemer}, includes the effect of periastron precession. One therefore measures the observable mean motion $\bn = \frac{2\pi}{\bar{P}} = n\left(1 + k\right)$, equation \eqref{eq:obsmeanmotion}. It follows that the so-called mass function from the pulsar timing (see e.g. Hobbs 2006) can be expressed as 
\begin{eqnarray}
\label{eq:massfunc}
f_m & = & \frac{(m_c\sin i)^3}{M^2} \\
& = & G^{-1}(a_p\sin i)^3\bn^2 \left( 1 - \delta -2k\right)
\end{eqnarray} 
where $a_p = am_c/M$ is the semi-major axis of the pulsar, $i$ is the inclination of the orbital plane relative to the plane of the sky, $k$ and $\delta$ are defined in \eqref{eq:k}-\eqref{eq:delta}. One can solve equation \eqref{eq:massfunc} for the companion mass $m_c$ if the pulsar projected semi-major axis $x = a_p\sin i$ is known (e.g. from the timing), and assumes/knows either the pulsar mass or the orbital inclination \citep[e.g. from optical light-curve modelling,][]{breton_discovery_2013}. At relevant order, one can solve for the uncorrected mass function (i.e. $\delta = k = 0$ above) and re-inject the obtained leading order solution $m_c^{(0)}$ to obtain the first order correction (the only one relevant here) by linearising \eqref{eq:massfunc}, thus giving explicitly
\begin{equation}
m_c = m_c^{(0)}\left( 1 - \frac{\rotqadim + 5\tidqadim + \epsilon\left(\frac{m_c^{(0)}m_p}{{3M^{(0)}}^2} + 1\right)}{1-\frac{2}{3}\frac{m_c^{(0)}}{M^{(0)}}}  \right),
\end{equation}
where $M^{(0)} = m_p + m_c^{(0)}$.

In spider binaries, the mass ratio $q$ can be measured directly through optical spectroscopy of the companion (see e.g. \citep{van_kerkwijk_evidence_2011}), leading to an estimate of the mass of the pulsar using the mass function \eqref{eq:massfunc}. Taking into account quadrupole and relativistic effects leads to a corrected pulsar mass,
\begin{equation}
\label{eq:mpcor}
m_p = m_p^{(0)}\left(1 - 3\rotqadim - 15\tidqadim - \epsilon\left(\frac{q}{(1+q)^2} + 3\right) \right),
\end{equation}
where $m_p^{(0)}$ is the uncorrected mass. 

As we can see, the obtained correction is of order $\sim k$ which is expected to be no more than $\sim 10^{-3}$ from equation \eqref{eq:J2final} and \eqref{eq:relcorrection}. In state-of-the-art measurements using optical observations of the companion (e.g. \citet{van_kerkwijk_evidence_2011,romani_spectroscopic_2015,linares_peering_2018}) this correction is largely dominated by uncertainties on the orbital inclination $i$, and to a smaller extent by the uncertainty on radial velocity measurements. In principle, such accuracy can be achieved using pulsar timing by measuring the Shapiro delay due to the companion (e.g. \citet{demorest_two-solar-mass_2010}), but so far no such measurement was possible for spiders, partly due to prolonged radio eclipses caused by evaporated material outflowing from the companion \citep[see, e.g.,][for a recent study]{polzin_low-frequency_2018}.  

\subsection{Minimum eccentricity, precession and spider age}
\label{sec:eccperiastronprec}
\newcommand{\omdottid}{\omdot_t}
\newcommand{\omdotspin}{\omdot_{s}}
\newcommand{\omdotrel}{\omdot_\mathrm{rel}}

\subsubsection{Minimum eccentricity}
The timing formula \eqref{eq:roemer} provides a measure of the DD eccentricity, the physical interpretation of which is important to differentiate from that of the osculating eccentricity defined in the Keplerian sense. In appendix \ref{secap:osculating} we show that the DD eccentricity $e$ is related to the osculating Keplerian eccentricity $e^K$ through equation \eqref{apeq:eKeDD}
\begin{equation}
\label{eq:eKemin}
	e = e^K + \emin,
\end{equation}
where the minimum eccentricity is defined as
\begin{equation}
\label{eq:emin}
	\emin = -(\delta + 2k),
\end{equation}
with $k$ and $\delta$ given in equations \eqref{eq:k} and \eqref{eq:delta}, respectively. Assuming that quadrupole precession dominates over relativistic precession, we can approximate using \eqref{eq:k} and \eqref{eq:omdottid}
\begin{equation}
\label{eq:eminest}
	\emin \simeq \omdottid/\bar{n} =  4.6\dix{-3}\left(\frac{k_2}{0.1} \right) \left(\frac{q}{\frac{1}{16}}\right)^{-1} f^5 \left(\frac{\rho_f(q)}{\rho_f(\frac{1}{16})}\right)^{5}.
\end{equation}

This minimum eccentricity can be interpreted as a parametrisation of the short-term perturbative effects presented in equations \eqref{apeq:dkc}-\eqref{apeq:dTa} rather than as a geometrical feature. Due to orbital precession, there is no solution which describes a closed ellipse. Had we chosen a different orbital parametrisation than the DD model (see appendix \ref{secap:pseudoDD}), extra terms of order $k$ would have been present in the Roemer delay (equation \eqref{eq:roemer}). This model would have been perfectly equivalent, though arguably more contrived.

Another conceptual way of interpreting the residual eccentricity is that the presence of extra terms at perturbation order in the trajectory (see \eqref{apeq:perturbtraj}) is an unavoidable consequence of Bertrand's theorem. This theorem states that strictly periodic trajectories can only be obtained from the $\propto 1/r$ Newtonian or the $\propto r^2$ harmonic oscillator potentials. We readily see here that a perfectly circular orbit which would only include secular precession as an effect of perturbations would be a violation of Bertrand's theorem as such a trajectory would be periodic and appear as circular under a redefinition of the orbital period.

We therefore conclude that while the osculating Keplerian eccentricity can be zero due to circularisation, an effective non-zero DD eccentricity component $e=\emin$ must remain and, should it be detected, can enable a measurement of $\dot{\omega}$.

\subsubsection{Eccentricity measurement and periastron precession}
\label{sec:ecc_precession}
To our knowledge eccentricity measurements have so far been reported for two spiders: the black widows PSR~J2051-0827 with $e\sim 5\dix{-5}$ \citep[see, e.g.,][]{lazaridis_evidence_2011,shaifullah_21_2016}) and PSR~J1731$-$1847 with $e\sim 3\dix{-5}$ \citep{ng_high_2014}. We note that an eccentricity measurement was originally reported for the redback PSR~J2339-0533 \citep{pletsch_gamma-ray_2015}, but has since been reconsidered to be consistent with zero (C. Clark, private communication).

In Section \ref{sec:constantqrelat} we showed that, together with relativistic effects, the spin and tidal quadrupole components are responsible for orbital precession and can in principle be measured due to the non-vanishing DD eccentricity. Using equations \eqref{eq:k} and \eqref{eq:omdot}  we can decompose the precession rate into three components, $\omdot = \omdotrel + \omdotspin + \omdottid$, which provide the relativistic, spin quadrupole and tidal quadrupole contributions, respectively: 
\begin{eqnarray}
	\omdotrel &=& 1.6\dix{-2}  \left(\frac{4\mathrm{h}}{P}\right)^{5/3} \left(\frac{M}{1.7\Msol}\right)^{2/3} \mathrm{cyc/yr},\\
	\omdotspin &=& -0.7 \left(\frac{4\mathrm{h}}{P}\right) \left(\frac{k_2}{0.1} \right) \left[ 1+ q\right] f^5 \left(\frac{\rho_f(q)}{\rho_f(\frac{1}{16})}\right)^{5} \mathrm{cyc/yr}, \\
	\label{eq:omdottid}
	\omdottid &=& -10 \left(\frac{4\mathrm{h}}{P}\right) \left(\frac{k_2}{0.1} \right) \left(\frac{q}{\frac{1}{16}}\right)^{-1} f^5 \left(\frac{\rho_f(q)}{\rho_f(\frac{1}{16})}\right)^{5} \mathrm{cyc/yr}.
\end{eqnarray}
From the above it is clear that the tidal contribution dominates.

We present estimates of $\omdot$ in Table \ref{tab:pulsars} for the considered subset of spider systems under both a large ($k_2=0.1$, $f = 0.9$) and a small ($k_2=0.01$, $f=0.5$) deformation hypothesis. Given the large predicted precession rates it appears that measuring the eccentricity with a timing model not accounting for this effect will average the value down and might, in several cases, wash it out entirely. It is quite likely that the lack of measured eccentricity so far in most spider systems is a consequence of the use of an improper timing model. We therefore strongly advocate switching to our proposed model to perform further spider timing.

One such example of a system that would benefit from our timing model is PSR~J2051$-$0827. The eccentricity is reported to vary between subsets of timing data covering a few years and attempts to measure it over the entire two decades of observations has failed to produce a consistent value \citep{lazaridis_evidence_2011,shaifullah_21_2016}. This behaviour is in line with the prediction from an unmodeled precession, and hence PSR~J2051$-$0827 is particularly promising to yield a measurement of $\dot{\omega}$. This, in turn, could open an unprecedented window into probing the internal structure of the companion by enabling us to constrain the apsidal motion constant $k_2$ and will be presented in a forthcoming publication.

The comparatively large minimum eccentricity implies that a (possibly large) fraction of timing residuals in some spider systems might be caused by precession. However, it is also possible for this signature to be small due to 1) the apsidal motion constant is small and thus the companion being quite `rigid' (e.g. more similar to low-mass white dwarf than a regular star; see Section \ref{sec:apsidalconstant}), and/or 2) the companion is significantly smaller than the Roche lobe. The tentative eccentricity measurements that have been reported do not allow to distinguish between these different cases as significant averaging might have occurred due to the use of unsuitable timing models.

\subsubsection{Spider age}
We expect spider binaries to be `born', e.g. emerging from the recycling process at the end of the low-mass X-ray binary phase, with an already very small Keplerian eccentricity $e^K_0 \ll 1$. From this point forward orbital eccentricity should be driven by tidal circularisation only -- this is likely an over-simplifying assumption --, and the osculating Keplerian eccentricity will then decay exponentially as $e^K(t) \sim e^K_0\exp(-t/\tau_c)$ \citep{hut_tidal_1981}, with the time scale given by equation \eqref{eq:circularisation}. One can therefore define an eccentricity age
\begin{equation}
\label{eq:taue}
	\tau_e = \tau_c \log\left(\frac{e^K_0}{e-\emin}\right),
\end{equation}
where $e$ is the DD eccentricity measured nowadays. We can then consider the upper limit $e^K_0=1$ and use the numbers from Section \ref{sec:tide} along with $e-\emin = 10^{-5}$ to obtain $\tau_e \lesssim 3\dix{8}$ and $3\dix{4}$yr for typical black widows and redbacks, respectively. Although these time scales are sensitive to the actual mass ratio and filling factor, the choice of $e_0$ is extremely conservative and hence the presence of any leftover Keplerian eccentricity should in principle be very unlikely. The detection of a spider with $e>\emin$ could indicate that the system was born recently or that a mechanism yet to identify pumps up orbital eccentricity in these systems \citep{ogilvie_tidal_2014}. It has been suggested that in pulsar-white dwarf binaries with observed eccentricities a circumbinary disc may temporarily form out of the material ejected by the companion after the end of the accretion phase and injects eccentricity into the system \citep{antoniadis_formation_2014}. The question is therefore open as to whether a similar mechanism could apply in spider systems.

\subsection{Effects of quadrupole variations}
\label{sec:discussvariations}
From Table \ref{tab:pulsars}, we see that the amplitude of the quadrupole variation,
\begin{equation}
\label{eq:normqadim}
	\normvqadim = \max(\vqadim(t)) - \min(\vqadim(t)),
\end{equation} 
is typically in the range $10^{-8} \lesssim \normvqadim \lesssim 10^{-7}$. The time scale $T_2$ of the variations is estimated as the leading harmonic of the Fourier transform of the period changes given by equation \eqref{eq:dp} or, in the case where less than one pseudo-period has been observed, by the period of the best sinusoidal fit. The observed variation time scale ranges from about 3 years to 22 years. We note three exceptions to the previous statement: PSRs~J0024-7204J, J1807-2459A and J1023+0038. In these three cases only one orbital frequency derivative could be observed and therefore the result of the fit may be unreliable. It is impossible to say, however, whether this situation arises from an observing time span which is not long enough or from the fact that the orbital period of these systems is intrinsically not oscillating. In the latter case it is more likely that a mechanism other than quadrupole variations is responsible (see below).

PSR~J1723-2837 displays by far the largest quadrupole amplitude, $\normvqadim \simeq 3\dix{-7}$, while also having the fastest period of variation, $T_2 \simeq 3$yr. Coincidentally, it is the only spider companion which shows evidence of asynchronous rotation from optical observations \citep{van_staden_active_2016}. This property might invalidate our primary assumption of phase-locking if the structure involved with asynchronous rotation extends beyond a thin shell of negligible mass. Whether this could in turn produce the seemingly exceptional values will require further investigation.

We should also remark even if the time scale of orbital period variations may be somewhat similar to that of constant-quadrupole precession, these two effects are not degenerate in a timing analysis. Indeed, the former effect is modulated at the orbital frequency, while the latter is modulated at twice that frequency, as can be seen from equation \eqref{eq:finalroemer}.

In Section \ref{sec:completeroemer} we made the assumption that short-term variations of the orbital elements can be neglected, $P/T_2 \ll 1$, in order to simplify the timing model. This assumption appears well justified as $P/T_2 \sim 10^{-4}$ for all systems presented in Table \ref{tab:pulsars}, meaning that extra-delays obtained without this approximation are of order $\sim P/T_2\Delta_R^{\sigma}\ll 1\mathrm{\mu s}$, which is smaller than the timing accuracy achievable in these systems. The same argument also justifies the approximations implicitly made in \citet{applegate_mechanism_1992}.

We note, however, a number of inconsistencies arising from existing work. Indeed, \citet{shaifullah_21_2016} report evidence that period variations may occur in PSR~J2051-0827 on shorter time scales, possibly less than 45 days. If these variations are related to aligned quadrupole variations, then the neglected short-term effects might eventually have to be reintroduced into the timing model. The same pulsar also displays important variations of its projected semi-major axis $x$ over time which are at odds with the magnitude with those expected from equation \eqref{eq:deltaRx}. We note that only one other pulsar has a reported variation of $x$, J0024-7204J, which is also at odds with the magnitude expected from period variations. In \citet{shaifullah_21_2016} it is also observed that the changes in $x$ seem uncorrelated with the changes in period, further suggesting that the two effects might have a different origin. They suggest that the mechanism could be spin-orbit coupling, proposing that a slight misalignment between the spin axis of the companion and the orbital angular momentum might produce the required changes in $x$. However they note that this misalignment should then change over time, a behaviour for which a mechanism is yet to be found. 

Beside the quadrupole, other effects such as mass loss due to evaporation and gravitational-wave radiation will induce changes in orbital period. It must be noted, however, that in both cases the effect is monotonous (the period increases and decreases respectively) and cannot explain the well-established oscillations of PSR~J2051-0827, for example. It has been shown in specific cases (particularly \citet{lazaridis_evidence_2011}) that the magnitude of these effects was too small by several orders of magnitude to explain the observed period variations. Another effect is magnetic braking (e.g. \citet{rappaport_new_1983, spruit_stellar_1983}), which also results in decreasing the orbital period by transferring angular momentum to the wind of the companion. However, fully convective stars such as spider companions \citep{chen_formation_2013} are thought to have a significantly reduced magnetic activity that nullifies magnetic braking (e.g. \citet{rappaport_new_1983, spruit_stellar_1983}). Finally, variations of the Doppler factor connecting the frame of the observer to the binary frame (Doppler delay in Section \ref{sec:timing}) can also cause apparent orbital period and semi-major axis variations. This can be estimated from position and proper motion measurements (see also  \citet{lazaridis_evidence_2011}). In all cases, if present, these effects are absorbed in the fit for orbital period derivatives (see equation \eqref{eq:nderivatives}).

\subsection{Timing of non-pulsar binary systems}
\label{sec:nonpulsar}
While we have focused most of this paper on spider binary pulsars, the timing model we propose can appropriately account for the orbital dynamics of a wider range of binary systems. Any such system must fulfill the basic requirements under which our mathematical development is valid, namely: 1) a small eccentricity, 2) a `primary' star that suffers negligible quadrupole distortion, and 3) a `secondary' star that is tidally locked, but which may suffer quadrupole distortions. Notable systems that fall under this category are hot Jupiters, hot subdwarf B (sdB) binaries or cataclysmic variables (CVs). It should be stressed that, even when no apsidal motion or eccentricity is detected, a simple upper limit on the eccentricity translates in an upper limit on the apsidal motion constant $k_2$ through the relation $e\geq\emin$. 

The structure of CV companions has been shown to be extremely centrally condensed, with apsidal motion constants as low as $k_2 \sim 10^{-4} -  10^{-2}$ \citep[][and Section \ref{sec:apsidalconstant}]{cisneros-parra_apsidal_1970, warner_apsidal_1978}. As a result, the expected apsidal motion as well as minimum eccentricity is often beyond reach of current observations \citep[e.g.][]{parsons_timing_2014}. Nonetheless, upper limits on eccentricity can still be turned into upper limits on $k_2$. For example, NN~Ser is an eclipsing system comprising a white dwarf and a post-common envelope secondary with excellent eclipse timing precision \citep{beuermann_two_2010, parsons_precise_2010, parsons_timing_2014}, but no detected eccentricity. The $0.1$s timing uncertainty of the primary eclipse translates in $\emin  < e \lesssim 2\dix{-5}$ and $k_2 \lesssim 0.005$, which falls within the expected range.

The 3-hour eclipsing sdB+M dwarf binary 2M~1938+4603 is reported to have an eccentricity $e \gtrsim 4 \dix{-5}$ and apsidal motion is not detected because of the short monitoring time span \citep{barlow_romer_2012}. Although the two components of the system are very different, it is not clear whether the quadrupole distortion of one star dominates over the other or if they are similar, in which case both contributions should blend  \citep[e.g.][]{kopal_dynamics_1978}. As this discussion is beyond the scope of the present paper, we only consider two extreme cases. Assuming the quadrupole moment of the M dwarf secondary dominates, the condition $e>\emin$ implies that $k_2 \lesssim 0.0025$ for the secondary. Such a low value is similar to what is expected for CV companions (see above) which is consistent with 2M~1938+4603 sharing a similar formation scenario with CVs. On the other hand, if the primary's quadrupole moment dominates, then its corresponding apsidal motion constant is $k_2 \lesssim 0.02$.

Hot Jupiters given share multiple similarities with black widows. A subset of these exoplanets have short, circular orbital periods ($\lesssim 1$\,d), fill a significant fraction of their Roche Lobe, are tidally locked, have an irradiated day side, and have a system mass ratio comparable to or more extreme than that of black widows \citep[see, e.g.,][and references therein for a recent review]{dawson_origins_2018}. Furthermore their apsidal motion constant is expected to be as high as $0.2$ \citep{kramm_constraining_2012}. Transit time variations would be expected in these systems due to precession induced by the minimum eccentricity. One challenge, however, is that this effect might be difficult to disentangle from gravitational perturbations caused by other planets in the system \citep{agol_detecting_2005} or rotating stellar spots on the host \citep{holczer_transit_2016}. We estimate the time scale of the transit time variations due to minimum eccentricity to range between sub-seconds to few minutes (assuming a fiducial Jupiter-mass planet in a 1-day orbit around a $1\,M_\odot$ star with $k_2=0.2$). This wide range of variations results from the extreme sensitivity on the assumed Roche lobe filling factor. While the typical precision achievable to measure transit times is currently of the order of minutes, more accurate timing should reveal that exoplanets must possess a residual minimum eccentricity.

\section{Conclusions}
 \label{sec:conclusions}
In this paper, we assumed that the non-Keplerian dynamical behaviour of spiders is dominated by the quadrupole moment of the companion star. This quadrupole can be decomposed into an intrinsic constant component due to centrifugal forces, an extrinsic component due to the equilibrium tides arising from the neutron star gravitational pull, and a variable component that is related to the companion's internal structure. Although the privileged mechanism for the latter is the so-called Applegate mechanism \citep{applegate_mechanism_1992, applegate_orbital_1994}, the treatment we propose here does not depend on a specific model. Instead, our approach focuses on finding a self-consistent solution to the dynamical response of the system to an arbitrary quadrupole perturbation. We also take into account relativistic effects -- essentially precession -- to first post-Newtonian order (1PN).

We show that in spider systems tidal interactions will rapidly phase-lock the companion's rotation and reduce the eccentricity to very small values. Both of these are the two main assumptions enabling us to find a solution to the orbital motion of a relativistic (1PN) binary, exact to first order in eccentricity, in which the companion possesses a tidal and centrifugal quadrupole moment. We call such a solution an effective DD solution, in reference to \cite{damour_general_1985} and \cite{wex_timing_1998}. We then add the effect of a variable quadrupole by matching Lagrange perturbation theory with our effective DD solution. Because the eccentricity is small, we adopt the Laplace-Lagrange parametrisation \citep{lange_precision_2001} of eccentricity and periastron longitude. We therefore obtain a generalisation of the ELL1 timing model \citep{lange_precision_2001} contained in the R\oe mer delay of equation \eqref{eq:finalroemer}. Our effective DD solution consistently accounts for precession, both quadrupole and relativistic, quadrupole variations, and low eccentricity. We believe that this new model should provide a much more appropriate parametrisation that can improve the timing of spider pulsars by accounting for additional systematic effects and possibly enable some of them to be incorporated into pulsar timing array experiments. This would represent a particularly important step change as a large fraction of the millisecond pulsar population is found in spider systems. However, we note that the prediction power of the model is diminished by the absence, to our knowledge, of a deterministic model for quadrupole variations. Predictability is however not necessary in pulsar timing arrays.

One particular application of our work is the possibility of measuring the tidal-centrifugal quadrupole moment of the companion star through the orbital precession it induces. Our precession model extends the one recently proposed in \cite{susobhanan_exploring_2018}: in addition to secular precession it accounts for short-term periodic effects caused by the tidal-centrifugal quadrupole and relativistic perturbations through a redefinition of the orbital elements summarized in Table \ref{tab:elements}. The main consequence of this new model is the existence of a so far unrecognised non-Keplerian minimum eccentricity, equation \ref{eq:eKemin}, which cannot be damped by circularisation effects. This eccentricity term scales with the quadrupole deformation of the companion and, given the large magnitude of tidal interactions existing in spider systems, we estimate (equation \eqref{eq:eminest} and Table \ref{tab:pulsars}) that is should be detectable in many if not all spider systems. This, in turn, implies that periastron precession is likely going to be measurable. We also show that orbital period variations and precession are uncorrelated effects since they are associated with different harmonics of the orbital period.

If detected, precession will shed a new light on the internal structure and evolution of these exotic stars. Together with optical observations to constrain the filling factor of the companion and the inclination of the system, a measurement of the quadrupole moment would give a direct constrain on the apsidal motion constant $k_2$, through equations \eqref{eq:rotqadim2}-\eqref{eq:tidqadim2}, which depends only on the internal structure of the star. We show that the induced precession could be very fast, possibly up to an entire orbit every few weeks, equation \eqref{eq:omdot}. 

We note that the fact that precession was so far not included in the timing models used for spiders might have dramatically hindered eccentricity measurements. Besides, the apparently variable eccentricity and periastron longitude reported for J2051-0827 \citep{shaifullah_21_2016} is strong evidence for precession. At the same time, we show that detecting larger-than-minimum eccentricities in some spider systems can hint to the fact that they are younger than expected. In an attempt to quantify this statement, we introduce the eccentricity age $\tau_e$ in equation \eqref{eq:taue} which gives the maximum time needed for the system to reach the observed eccentricity after the last mass transfer episode. For instance, a black widow with an eccentricity $e-\emin\sim 10^{-5}$ cannot have left mass transfer for more than a few hundred million years (or a few ten million years in the case of a redback).

Additionally, we provide explicit formulae to connect observables with physical quantities, especially for mass measurements, equations \eqref{eq:massfunc}-\eqref{eq:mpcor} (see also appendix \ref{secap:osculating}). In particular, we show that periastron precession can change the mass measurements by up to $0.1\%$ meaning that this effect should be treated as an additional systematic uncertainty. For the moment though, the error budget is dominated by other sources of errors such as the precision with which the orbital inclination is derived from optical data (e.g. \citet{linares_peering_2018}).

We showed in Section \ref{sec:discussvariations} that the effects of quadrupole variations on a long time scale can be accounted for by our model, while those occurring on a scale commensurate to the orbital period can be safely ignored. On the other hand, we note that a number of intriguing observations in PSRs~J2051-0827 and J0024-7204J such as large projected semi-major axis variations cannot be accounted for by a time-variable, aligned quadrupole (see Table \ref{tab:pulsars}). As suggested in \citet{lazaridis_evidence_2011}, these might require the inclusion of further physical effects such as spin-orbit coupling.

We also highlight that the tidal interaction induces a low-frequency timing delay at the precession frequency, which could otherwise be absorbed into the red noise typically present in timing residuals. Our timing model can properly disentangle this effect.

Finally, we demonstrate that our orbital solution extends the framework already established in other areas of binary stellar astrophysics. Our prediction of a minimum eccentricity allows one to infer an upper limit on the apsidal motion constant even in the absence of a measurable orbital precession.

\section*{Acknowledgements}

The authors acknowledge support of the European Research Council, under the European Union's Horizon 2020 research and innovation programme (grant agreement No. 715051; Spiders). This work made use of SageMath (\url{http://www.sagemath.org}). The authors wish to thank Mark Kennedy, Vik Dhillon, Colin Clark and Daniel Mata S\'anchez for insightful discussions about the manuscript.




\bibliographystyle{mnras}
\bibliography{biblio} 

\begin{thebibliography}{}
\makeatletter
\relax
\def\mn@urlcharsother{\let\do\@makeother \do\$\do\&\do\#\do\^\do\_\do\%\do\~}
\def\mn@doi{\begingroup\mn@urlcharsother \@ifnextchar [ {\mn@doi@}
  {\mn@doi@[]}}
\def\mn@doi@[#1]#2{\def\@tempa{#1}\ifx\@tempa\@empty \href
  {http://dx.doi.org/#2} {doi:#2}\else \href {http://dx.doi.org/#2} {#1}\fi
  \endgroup}
\def\mn@eprint#1#2{\mn@eprint@#1:#2::\@nil}
\def\mn@eprint@arXiv#1{\href {http://arxiv.org/abs/#1} {{\tt arXiv:#1}}}
\def\mn@eprint@dblp#1{\href {http://dblp.uni-trier.de/rec/bibtex/#1.xml}
  {dblp:#1}}
\def\mn@eprint@#1:#2:#3:#4\@nil{\def\@tempa {#1}\def\@tempb {#2}\def\@tempc
  {#3}\ifx \@tempc \@empty \let \@tempc \@tempb \let \@tempb \@tempa \fi \ifx
  \@tempb \@empty \def\@tempb {arXiv}\fi \@ifundefined
  {mn@eprint@\@tempb}{\@tempb:\@tempc}{\expandafter \expandafter \csname
  mn@eprint@\@tempb\endcsname \expandafter{\@tempc}}}

\bibitem[\protect\citeauthoryear{Agol, Steffen, Sari  \& Clarkson}{Agol
  et~al.}{2005}]{agol_detecting_2005}
Agol E.,  Steffen J.,  Sari R.,   Clarkson W.,  2005, \mn@doi [Monthly Notices
  of the Royal Astronomical Society] {10.1111/j.1365-2966.2005.08922.x}, 359,
  567

\bibitem[\protect\citeauthoryear{Alpar, Cheng, Ruderman  \& Shaham}{Alpar
  et~al.}{1982}]{alpar_new_1982}
Alpar M.~A.,  Cheng A.~F.,  Ruderman M.~A.,   Shaham J.,  1982, \mn@doi
  [Nature] {10.1038/300728a0}, 300, 728

\bibitem[\protect\citeauthoryear{Antoniadis}{Antoniadis}{2014}]{antoniadis_formation_2014}
Antoniadis J.,  2014, \mn@doi [The Astrophysical Journal]
  {10.1088/2041-8205/797/2/L24}, 797, L24

\bibitem[\protect\citeauthoryear{Antoniadis, Tauris, Ozel, Barr, Champion  \&
  Freire}{Antoniadis et~al.}{2016}]{antoniadis_millisecond_2016}
Antoniadis J.,  Tauris T.~M.,  Ozel F.,  Barr E.,  Champion D.~J.,   Freire P.
  C.~C.,  2016, preprint, p. arXiv:1605.01665

\bibitem[\protect\citeauthoryear{Applegate}{Applegate}{1992}]{applegate_mechanism_1992}
Applegate J.~H.,  1992, \mn@doi [The Astrophysical Journal] {10.1086/170967},
  385, 621

\bibitem[\protect\citeauthoryear{Applegate \& Shaham}{Applegate \&
  Shaham}{1994}]{applegate_orbital_1994}
Applegate J.~H.,  Shaham J.,  1994, \mn@doi [The Astrophysical Journal]
  {10.1086/174906}, 436, 312

\bibitem[\protect\citeauthoryear{Archibald et~al.,}{Archibald
  et~al.}{2009}]{archibald_radio_2009}
Archibald A.~M.,  et~al., 2009, \mn@doi [Science] {10.1126/science.1172740},
  324, 1411

\bibitem[\protect\citeauthoryear{Arzoumanian, Fruchter  \& Taylor}{Arzoumanian
  et~al.}{1994}]{arzoumanian_orbital_1994}
Arzoumanian Z.,  Fruchter A.~S.,   Taylor J.~H.,  1994, \mn@doi [The
  Astrophysical Journal] {10.1086/187346}, 426, L85

\bibitem[\protect\citeauthoryear{Barlow, Wade  \& Liss}{Barlow
  et~al.}{2012}]{barlow_romer_2012}
Barlow B.~N.,  Wade R.~A.,   Liss S.~E.,  2012, \mn@doi [The Astrophysical
  Journal] {10.1088/0004-637X/753/2/101}, 753, 101

\bibitem[\protect\citeauthoryear{Bassa et~al.,}{Bassa
  et~al.}{2014}]{bassa_state_2014}
Bassa C.~G.,  et~al., 2014, \mn@doi [Monthly Notices of the Royal Astronomical
  Society] {10.1093/mnras/stu708}, 441, 1825

\bibitem[\protect\citeauthoryear{Bassa et~al.,}{Bassa
  et~al.}{2017}]{bassa_lofar_2017}
Bassa C.~G.,  et~al., 2017, \mn@doi [The Astrophysical Journal]
  {10.3847/2041-8213/aa8400}, 846, L20

\bibitem[\protect\citeauthoryear{Bellm et~al.,}{Bellm
  et~al.}{2016}]{bellm_properties_2016}
Bellm E.~C.,  et~al., 2016, \mn@doi [The Astrophysical Journal]
  {10.3847/0004-637X/816/2/74}, 816, 74

\bibitem[\protect\citeauthoryear{Benvenuto, De~Vito  \& Horvath}{Benvenuto
  et~al.}{2012}]{benvenuto_evolutionary_2012}
Benvenuto O.~G.,  De~Vito M.~A.,   Horvath J.~E.,  2012, \mn@doi [The
  Astrophysical Journal] {10.1088/2041-8205/753/2/L33}, 753, L33

\bibitem[\protect\citeauthoryear{Beuermann et~al.,}{Beuermann
  et~al.}{2010}]{beuermann_two_2010}
Beuermann K.,  et~al., 2010, \mn@doi [Astronomy and Astrophysics]
  {10.1051/0004-6361/201015472}, 521, L60

\bibitem[\protect\citeauthoryear{Beutler}{Beutler}{2004}]{beutler_methods_2004}
Beutler G.,  2004, Methods of {Celestial} {Mechanics}: {Volume} {I}:
  {Physical}, {Mathematical}, and {Numerical} {Principles}, 1st softcover
  edition without cd-rom of original hardcover edition. edition edn.
Springer, Berlin ; New York

\bibitem[\protect\citeauthoryear{Blandford \& Teukolsky}{Blandford \&
  Teukolsky}{1976}]{blandford_arrival-time_1976}
Blandford R.,  Teukolsky S.~A.,  1976, \mn@doi [The Astrophysical Journal]
  {10.1086/154315}, 205, 580

\bibitem[\protect\citeauthoryear{Breton et~al.,}{Breton
  et~al.}{2013}]{breton_discovery_2013}
Breton R.~P.,  et~al., 2013, \mn@doi [The Astrophysical Journal]
  {10.1088/0004-637X/769/2/108}, 769, 108

\bibitem[\protect\citeauthoryear{Brinkworth, Marsh, Dhillon  \&
  Knigge}{Brinkworth et~al.}{2006}]{brinkworth_detection_2006}
Brinkworth C.~S.,  Marsh T.~R.,  Dhillon V.~S.,   Knigge C.,  2006, \mn@doi
  [Monthly Notices of the Royal Astronomical Society]
  {10.1111/j.1365-2966.2005.09718.x}, 365, 287

\bibitem[\protect\citeauthoryear{Burderi, D'Antona, Di~Salvo  \&
  Burgay}{Burderi et~al.}{2002}]{burderi_radio_2002}
Burderi L.,  D'Antona F.,  Di~Salvo T.,   Burgay M.,  2002,
  arXiv:astro-ph/0208021

\bibitem[\protect\citeauthoryear{Chen, Chen, Tauris  \& Han}{Chen
  et~al.}{2013}]{chen_formation_2013}
Chen H.-L.,  Chen X.,  Tauris T.~M.,   Han Z.,  2013, \mn@doi [The
  Astrophysical Journal] {10.1088/0004-637X/775/1/27}, 775, 27

\bibitem[\protect\citeauthoryear{Cisneros-Parra}{Cisneros-Parra}{1970}]{cisneros-parra_apsidal_1970}
Cisneros-Parra J.~U.,  1970, Astronomy and Astrophysics, 8, 141

\bibitem[\protect\citeauthoryear{Coles, Hobbs, Champion, Manchester  \&
  Verbiest}{Coles et~al.}{2011}]{coles_pulsar_2011}
Coles W.,  Hobbs G.,  Champion D.~J.,  Manchester R.~N.,   Verbiest J. P.~W.,
  2011, \mn@doi [Monthly Notices of the Royal Astronomical Society]
  {10.1111/j.1365-2966.2011.19505.x}, 418, 561

\bibitem[\protect\citeauthoryear{Crawford et~al.,}{Crawford
  et~al.}{2013}]{crawford_psr_2013}
Crawford F.,  et~al., 2013, \mn@doi [The Astrophysical Journal]
  {10.1088/0004-637X/776/1/20}, 776, 20

\bibitem[\protect\citeauthoryear{Damour \& Deruelle}{Damour \&
  Deruelle}{1985}]{damour_general_1985}
Damour T.,  Deruelle N.,  1985, Ann. Inst. Henri Poincar{\'e} Phys. Th{\'e}or.,
  Vol. 43, No. 1, p. 107 - 132, 43, 107

\bibitem[\protect\citeauthoryear{Damour \& Deruelle}{Damour \&
  Deruelle}{1986}]{damour_general_1986}
Damour T.,  Deruelle N.,  1986, Annales de l'institut Henri Poincar{\'e} (A)
  Physique th{\'e}orique, 44, 263

\bibitem[\protect\citeauthoryear{Damour \& Taylor}{Damour \&
  Taylor}{1992}]{damour_strong-field_1992}
Damour T.,  Taylor J.~H.,  1992, \mn@doi [Physical Review D]
  {10.1103/PhysRevD.45.1840}, 45, 1840

\bibitem[\protect\citeauthoryear{Dawson \& Johnson}{Dawson \&
  Johnson}{2018}]{dawson_origins_2018}
Dawson R.~I.,  Johnson J.~A.,  2018, \mn@doi [Annual Review of Astronomy and
  Astrophysics] {10.1146/annurev-astro-081817-051853}, 56, 175

\bibitem[\protect\citeauthoryear{Demorest, Pennucci, Ransom, Roberts  \&
  Hessels}{Demorest et~al.}{2010}]{demorest_two-solar-mass_2010}
Demorest P.~B.,  Pennucci T.,  Ransom S.~M.,  Roberts M. S.~E.,   Hessels J.
  W.~T.,  2010, \mn@doi [Nature] {10.1038/nature09466}, 467, 1081

\bibitem[\protect\citeauthoryear{Edwards, Hobbs  \& Manchester}{Edwards
  et~al.}{2006}]{edwards_tempo2_2006}
Edwards R.~T.,  Hobbs G.~B.,   Manchester R.~N.,  2006, \mn@doi [Monthly
  Notices of the Royal Astronomical Society]
  {10.1111/j.1365-2966.2006.10870.x}, 372, 1549

\bibitem[\protect\citeauthoryear{Eggleton}{Eggleton}{1983}]{eggleton_aproximations_1983}
Eggleton P.~P.,  1983, \mn@doi [The Astrophysical Journal] {10.1086/160960},
  268, 368

\bibitem[\protect\citeauthoryear{Freire, Camilo, Kramer, Lorimer, Lyne,
  Manchester  \& D'Amico}{Freire et~al.}{2003}]{freire_further_2003}
Freire P.~C.,  Camilo F.,  Kramer M.,  Lorimer D.~R.,  Lyne A.~G.,  Manchester
  R.~N.,   D'Amico N.,  2003, \mn@doi [Monthly Notices of the Royal
  Astronomical Society] {10.1046/j.1365-8711.2003.06392.x}, 340, 1359

\bibitem[\protect\citeauthoryear{Fruchter, Stinebring  \& Taylor}{Fruchter
  et~al.}{1988}]{fruchter_millisecond_1988}
Fruchter A.~S.,  Stinebring D.~R.,   Taylor J.~H.,  1988, \mn@doi [Nature]
  {10.1038/333237a0}, 333, 237

\bibitem[\protect\citeauthoryear{{Hern{\'a}ndez Santisteban}
  et~al.,}{{Hern{\'a}ndez Santisteban}
  et~al.}{2016}]{hernandez_irradiated_2016}
{Hern{\'a}ndez Santisteban} J.~V.,  et~al., 2016, \mn@doi [\nat]
  {10.1038/nature17952}, \href
  {https://ui.adsabs.harvard.edu/abs/2016Natur.533..366H} {533, 366}

\bibitem[\protect\citeauthoryear{Hessels, Ransom, Stairs, Freire, Kaspi  \&
  Camilo}{Hessels et~al.}{2006}]{hessels_radio_2006}
Hessels J. W.~T.,  Ransom S.~M.,  Stairs I.~H.,  Freire P. C.~C.,  Kaspi V.~M.,
    Camilo F.,  2006, \mn@doi [Science] {10.1126/science.1123430}, 311, 1901

\bibitem[\protect\citeauthoryear{Hobbs, Edwards  \& Manchester}{Hobbs
  et~al.}{2006}]{hobbs_tempo2_2006}
Hobbs G.~B.,  Edwards R.~T.,   Manchester R.~N.,  2006, \mn@doi [Monthly
  Notices of the Royal Astronomical Society]
  {10.1111/j.1365-2966.2006.10302.x}, 369, 655

\bibitem[\protect\citeauthoryear{Holczer et~al.,}{Holczer
  et~al.}{2016}]{holczer_transit_2016}
Holczer T.,  et~al., 2016, \mn@doi [The Astrophysical Journal Supplement
  Series] {10.3847/0067-0049/225/1/9}, 225, 9

\bibitem[\protect\citeauthoryear{Hurley, Tout  \& Pols}{Hurley
  et~al.}{2002}]{hurley_evolution_2002}
Hurley J.~R.,  Tout C.~A.,   Pols O.~R.,  2002, \mn@doi [Monthly Notices of the
  Royal Astronomical Society] {10.1046/j.1365-8711.2002.05038.x}, 329, 897

\bibitem[\protect\citeauthoryear{Hut}{Hut}{1981}]{hut_tidal_1981}
Hut P.,  1981, Astronomy and Astrophysics, 99, 126

\bibitem[\protect\citeauthoryear{Kennedy, Clark, Voisin  \& Breton}{Kennedy
  et~al.}{2018}]{kennedy_kepler_2018}
Kennedy M.~R.,  Clark C.~J.,  Voisin G.,   Breton R.~P.,  2018,
  arXiv:1801.10609 [astro-ph]

\bibitem[\protect\citeauthoryear{Kopal}{Kopal}{1978}]{kopal_dynamics_1978}
Kopal Z.,  1978, Dynamics of {Close} {Binary} {Systems}.
Springer Netherlands, Dordrecht, \url
  {http://dx.doi.org/10.1007/978-94-009-9780-6}

\bibitem[\protect\citeauthoryear{Kopeikin}{Kopeikin}{1995}]{kopeikin_possible_1995}
Kopeikin S.~M.,  1995, \mn@doi [The Astrophysical Journal Letters]
  {10.1086/187731}, 439, L5

\bibitem[\protect\citeauthoryear{Kopeikin}{Kopeikin}{1996}]{kopeikin_proper_1996}
Kopeikin S.~M.,  1996, \mn@doi [The Astrophysical Journal Letters]
  {10.1086/310201}, 467, L93

\bibitem[\protect\citeauthoryear{Kramm, Nettelmann, Redmer  \& Stevenson}{Kramm
  et~al.}{2011}]{kramm_degeneracy_2011}
Kramm U.,  Nettelmann N.,  Redmer R.,   Stevenson D.~J.,  2011, \mn@doi
  [Astronomy and Astrophysics] {10.1051/0004-6361/201015803}, 528, A18

\bibitem[\protect\citeauthoryear{Kramm, Nettelmann, Fortney, Neuh{\"a}user  \&
  Redmer}{Kramm et~al.}{2012}]{kramm_constraining_2012}
Kramm U.,  Nettelmann N.,  Fortney J.~J.,  Neuh{\"a}user R.,   Redmer R.,
  2012, \mn@doi [Astronomy and Astrophysics] {10.1051/0004-6361/201118141},
  538, A146

\bibitem[\protect\citeauthoryear{Lange, Camilo, Wex, Kramer, Backer, Lyne  \&
  Doroshenko}{Lange et~al.}{2001}]{lange_precision_2001}
Lange C.,  Camilo F.,  Wex N.,  Kramer M.,  Backer D.~C.,  Lyne A.~G.,
  Doroshenko O.,  2001, \mn@doi [Monthly Notices of the Royal Astronomical
  Society] {10.1046/j.1365-8711.2001.04606.x}, 326, 274

\bibitem[\protect\citeauthoryear{Lanza}{Lanza}{2005}]{lanza_orbital_2005}
Lanza A.~F.,  2005, \mn@doi [Monthly Notices of the Royal Astronomical Society]
  {10.1111/j.1365-2966.2005.09559.x}, 364, 238

\bibitem[\protect\citeauthoryear{Lanza}{Lanza}{2006}]{lanza_time_2006}
Lanza A.~F.,  2006, \mn@doi [Monthly Notices of the Royal Astronomical Society]
  {10.1111/j.1365-2966.2006.11085.x}, 373, 819

\bibitem[\protect\citeauthoryear{Lanza \& Rodon{\`o}}{Lanza \&
  Rodon{\`o}}{1999}]{lanza_orbital_1999}
Lanza A.~F.,  Rodon{\`o} M.,  1999, Astronomy and Astrophysics, 349, 887

\bibitem[\protect\citeauthoryear{Lanza, Rodono  \& Rosner}{Lanza
  et~al.}{1998}]{lanza_orbital_1998}
Lanza A.~F.,  Rodono M.,   Rosner R.,  1998, \mn@doi [Monthly Notices of the
  Royal Astronomical Society] {10.1046/j.1365-8711.1998.01446.x}, 296, 893

\bibitem[\protect\citeauthoryear{Lazaridis et~al.,}{Lazaridis
  et~al.}{2011}]{lazaridis_evidence_2011}
Lazaridis K.,  et~al., 2011, \mn@doi [Monthly Notices of the Royal Astronomical
  Society] {10.1111/j.1365-2966.2011.18610.x}, 414, 3134

\bibitem[\protect\citeauthoryear{Linares, Shahbaz  \& Casares}{Linares
  et~al.}{2018}]{linares_peering_2018}
Linares M.,  Shahbaz T.,   Casares J.,  2018, \mn@doi [The Astrophysical
  Journal] {10.3847/1538-4357/aabde6}, 859, 54

\bibitem[\protect\citeauthoryear{Lynch, Freire, Ransom  \& Jacoby}{Lynch
  et~al.}{2012}]{lynch_timing_2012}
Lynch R.~S.,  Freire P. C.~C.,  Ransom S.~M.,   Jacoby B.~A.,  2012, \mn@doi
  [The Astrophysical Journal] {10.1088/0004-637X/745/2/109}, 745, 109

\bibitem[\protect\citeauthoryear{Manchester}{Manchester}{2017}]{manchester_millisecond_2017}
Manchester R.~N.,  2017, \mn@doi [Journal of Astrophysics and Astronomy]
  {10.1007/s12036-017-9469-2}, 38, 42

\bibitem[\protect\citeauthoryear{Marshall, Guillemot, Harding, Martin  \&
  Smith}{Marshall et~al.}{2016}]{marshall_new_2016}
Marshall F.~E.,  Guillemot L.,  Harding A.~K.,  Martin P.,   Smith D.~A.,
  2016, \mn@doi [The Astrophysical Journal Letters]
  {10.3847/2041-8205/827/2/L39}, 827, L39

\bibitem[\protect\citeauthoryear{Navarrete, Schleicher, K{\"a}pyl{\"a},
  Schober, V{\"o}lschow  \& Mennickent}{Navarrete
  et~al.}{}]{navarrete_magneto-hydrodynamical_nodate}
Navarrete F.~H.,  Schleicher D. R.~G.,  K{\"a}pyl{\"a} P.~J.,  Schober J.,
  V{\"o}lschow M.,   Mennickent R.~E., , \mn@doi [Monthly Notices of the Royal
  Astronomical Society] {10.1093/mnras/stz3065}

\bibitem[\protect\citeauthoryear{Ng et~al.,}{Ng et~al.}{2014}]{ng_high_2014}
Ng C.,  et~al., 2014, \mn@doi [Monthly Notices of the Royal Astronomical
  Society] {10.1093/mnras/stu067}, 439, 1865

\bibitem[\protect\citeauthoryear{Ogilvie}{Ogilvie}{2014}]{ogilvie_tidal_2014}
Ogilvie G.~I.,  2014, \mn@doi [Annual Review of Astronomy and Astrophysics]
  {10.1146/annurev-astro-081913-035941}, 52, 171

\bibitem[\protect\citeauthoryear{{\"O}zel \& Freire}{{\"O}zel \&
  Freire}{2016}]{ozel_masses_2016}
{\"O}zel F.,  Freire P.,  2016, \mn@doi [Annual Review of Astronomy and
  Astrophysics] {10.1146/annurev-astro-081915-023322}, 54, 401

\bibitem[\protect\citeauthoryear{Papitto et~al.,}{Papitto
  et~al.}{2013}]{papitto_swings_2013}
Papitto A.,  et~al., 2013, \mn@doi [Nature] {10.1038/nature12470}, 501, 517

\bibitem[\protect\citeauthoryear{Parsons, Marsh, Copperwheat, Dhillon,
  Littlefair, G{\"a}nsicke  \& Hickman}{Parsons
  et~al.}{2010}]{parsons_precise_2010}
Parsons S.~G.,  Marsh T.~R.,  Copperwheat C.~M.,  Dhillon V.~S.,  Littlefair
  S.~P.,  G{\"a}nsicke B.~T.,   Hickman R.,  2010, \mn@doi [Monthly Notices of
  the Royal Astronomical Society] {10.1111/j.1365-2966.2009.16072.x}, 402, 2591

\bibitem[\protect\citeauthoryear{Parsons et~al.,}{Parsons
  et~al.}{2014}]{parsons_timing_2014}
Parsons S.~G.,  et~al., 2014, \mn@doi [Monthly Notices of the Royal
  Astronomical Society] {10.1093/mnrasl/slt169}, 438, L91

\bibitem[\protect\citeauthoryear{Pletsch \& Clark}{Pletsch \&
  Clark}{2015}]{pletsch_gamma-ray_2015}
Pletsch H.~J.,  Clark C.~J.,  2015, \mn@doi [The Astrophysical Journal]
  {10.1088/0004-637X/807/1/18}, 807, 18

\bibitem[\protect\citeauthoryear{Podsiadlowski}{Podsiadlowski}{1991}]{podsiadlowski_irradiation-driven_1991}
Podsiadlowski P.,  1991, \mn@doi [Nature] {10.1038/350136a0}, 350, 136

\bibitem[\protect\citeauthoryear{Podsiadlowski, Rappaport  \&
  Pfahl}{Podsiadlowski et~al.}{2002}]{podsiadlowski_evolutionary_2002}
Podsiadlowski P.,  Rappaport S.,   Pfahl E.~D.,  2002, \mn@doi [The
  Astrophysical Journal] {10.1086/324686}, 565, 1107

\bibitem[\protect\citeauthoryear{Polzin et~al.,}{Polzin
  et~al.}{2018}]{polzin_low-frequency_2018}
Polzin E.~J.,  et~al., 2018, \mn@doi [Monthly Notices of the Royal Astronomical
  Society] {10.1093/mnras/sty349}, 476, 1968

\bibitem[\protect\citeauthoryear{Rappaport, Verbunt  \& Joss}{Rappaport
  et~al.}{1983}]{rappaport_new_1983}
Rappaport S.,  Verbunt F.,   Joss P.~C.,  1983, \mn@doi [The Astrophysical
  Journal] {10.1086/161569}, 275, 713

\bibitem[\protect\citeauthoryear{Rasio, Tout, Lubow  \& Livio}{Rasio
  et~al.}{1996}]{rasio_tidal_1996}
Rasio F.~A.,  Tout C.~A.,  Lubow S.~H.,   Livio M.,  1996, \mn@doi [The
  Astrophysical Journal] {10.1086/177941}, 470, 1187

\bibitem[\protect\citeauthoryear{Roberts}{Roberts}{2012}]{roberts_surrounded_2012}
Roberts M.~S.,  2012, Proceedings of the International Astronomical Union, 8,
  127

\bibitem[\protect\citeauthoryear{Romani, Filippenko  \& Cenko}{Romani
  et~al.}{2015}]{romani_spectroscopic_2015}
Romani R.~W.,  Filippenko A.~V.,   Cenko S.~B.,  2015, \mn@doi [The
  Astrophysical Journal] {10.1088/0004-637X/804/2/115}, 804, 115

\bibitem[\protect\citeauthoryear{Shaifullah et~al.,}{Shaifullah
  et~al.}{2016}]{shaifullah_21_2016}
Shaifullah G.,  et~al., 2016, \mn@doi [Monthly Notices of the Royal
  Astronomical Society] {10.1093/mnras/stw1737}, 462, 1029

\bibitem[\protect\citeauthoryear{Spruit \& Ritter}{Spruit \&
  Ritter}{1983}]{spruit_stellar_1983}
Spruit H.~C.,  Ritter H.,  1983, Astronomy and Astrophysics, 124, 267

\bibitem[\protect\citeauthoryear{Stappers, Bailes, Lyne, Camilo, Manchester,
  Sandhu, Toscano  \& Bell}{Stappers et~al.}{2001}]{stappers_nature_2001}
Stappers B.~W.,  Bailes M.,  Lyne A.~G.,  Camilo F.,  Manchester R.~N.,  Sandhu
  J.~S.,  Toscano M.,   Bell J.~F.,  2001, \mn@doi [Monthly Notices of the
  Royal Astronomical Society] {10.1046/j.1365-8711.2001.04074.x}, 321, 576

\bibitem[\protect\citeauthoryear{Sterne}{Sterne}{1939}]{sterne_apsidal_1939}
Sterne T.~E.,  1939, \mn@doi [Monthly Notices of the Royal Astronomical
  Society] {10.1093/mnras/99.5.451}, 99, 451

\bibitem[\protect\citeauthoryear{Susobhanan, Gopakumar, Joshi  \&
  Kumar}{Susobhanan et~al.}{2018}]{susobhanan_exploring_2018}
Susobhanan A.,  Gopakumar A.,  Joshi B.~C.,   Kumar R.,  2018, \mn@doi [Monthly
  Notices of the Royal Astronomical Society] {10.1093/mnras/sty2177}, 480, 5260

\bibitem[\protect\citeauthoryear{Tauris et~al.,}{Tauris
  et~al.}{2017}]{tauris_formation_2017}
Tauris T.~M.,  et~al., 2017, \mn@doi [The Astrophysical Journal]
  {10.3847/1538-4357/aa7e89}, 846, 170

\bibitem[\protect\citeauthoryear{Valsecchi, Farr, Willems, Deloye  \&
  Kalogera}{Valsecchi et~al.}{2012}]{valsecchi_tidally_2012}
Valsecchi F.,  Farr W.~M.,  Willems B.,  Deloye C.~J.,   Kalogera V.,  2012,
  \mn@doi [The Astrophysical Journal] {10.1088/0004-637X/745/2/137}, 745, 137

\bibitem[\protect\citeauthoryear{Voisin}{Voisin}{2017}]{voisin_simulation_2017}
Voisin G.,  2017, Theses, Universit{\'e} de recherche Paris Sciences et
  Lettres, \url {https://hal.archives-ouvertes.fr/tel-01677325}

\bibitem[\protect\citeauthoryear{V{\"o}lschow, Schleicher, Perdelwitz  \&
  Banerjee}{V{\"o}lschow et~al.}{2016}]{volschow_eclipsing_2016}
V{\"o}lschow M.,  Schleicher D. R.~G.,  Perdelwitz V.,   Banerjee R.,  2016,
  \mn@doi [Astronomy \& Astrophysics] {10.1051/0004-6361/201527333}, 587, A34

\bibitem[\protect\citeauthoryear{V{\"o}lschow, Schleicher, Banerjee  \&
  Schmitt}{V{\"o}lschow et~al.}{2018}]{volschow_physics_2018}
V{\"o}lschow M.,  Schleicher D. R.~G.,  Banerjee R.,   Schmitt J. H. M.~M.,
  2018, \mn@doi [Astronomy and Astrophysics] {10.1051/0004-6361/201833506},
  620, A42

\bibitem[\protect\citeauthoryear{Warner}{Warner}{1978}]{warner_apsidal_1978}
Warner B.,  1978, Acta Astronomica, 28, 303

\bibitem[\protect\citeauthoryear{Wex}{Wex}{1998}]{wex_timing_1998}
Wex N.,  1998, \mn@doi [Monthly Notices of the Royal Astronomical Society]
  {10.1046/j.1365-8711.1998.01570.x}, 298, 67

\bibitem[\protect\citeauthoryear{Zahn}{Zahn}{1977}]{zahn_reprint_1977}
Zahn J.-P.,  1977, Astronomy and Astrophysics, 500, 121

\bibitem[\protect\citeauthoryear{van Haasteren \& Levin}{van Haasteren \&
  Levin}{2013}]{van_haasteren_understanding_2013}
van Haasteren R.,  Levin Y.,  2013, \mn@doi [Monthly Notices of the Royal
  Astronomical Society] {10.1093/mnras/sts097}, 428, 1147

\bibitem[\protect\citeauthoryear{van Kerkwijk, Breton  \& Kulkarni}{van
  Kerkwijk et~al.}{2011}]{van_kerkwijk_evidence_2011}
van Kerkwijk M.~H.,  Breton R.~P.,   Kulkarni S.~R.,  2011, \mn@doi [The
  Astrophysical Journal] {10.1088/0004-637X/728/2/95}, 728, 95

\bibitem[\protect\citeauthoryear{van Staden \& Antoniadis}{van Staden \&
  Antoniadis}{2016}]{van_staden_active_2016}
van Staden A.~D.,  Antoniadis J.,  2016, \mn@doi [The Astrophysical Journal]
  {10.3847/2041-8213/833/1/L12}, 833, L12

\makeatother
\end{thebibliography}




\appendix

\section{Osculating orbit to the DD solution}
\label{secap:osculating}

The correspondence between the observable orbital elements, denoted with a bar, and the DD elements has been explicited in Sections \ref{sec:constantqrelat} and \ref{sec:timing}. We collate them in Table \ref{tab:elements}. Additionally we work out below the relationship between the elements describing the osculating Keplerian orbit to the DD solution at the time of periastron passage. We choose this particular epoch because of the simplicity the relationship has then. 

\newcommand{\laplace}{L}
The osculating eccentricity $e^{K}$ and semi-latus rectum $p^{K}$ are obtained from the Laplace vector integral $\vec{\laplace}$ and the orbital angular momentum $\vec{h}$ (see e.g. \citet{beutler_methods_2004}),
\begin{eqnarray}
	p^{K} & = & \frac{h^2}{GM}, \\
	e^K & = & \frac{q}{GM},
\end{eqnarray}
where 
\begin{eqnarray}
	\vec{h} & = & \vr\times\vrdot, \\
	\vec{\laplace} & = & \vrdot \times \vec{h} - GM \frac{\vr}{r}.
\end{eqnarray}
Expressing $\vr$ and $\vrdot$ at $t=T_p$ using the DD equations \eqref{eq:DDpos}-\eqref{eq:DD3rdlaw} and expanding to first order in eccentricity and perturbation one gets
\begin{equation}
\label{apeq:eKeDD}
	e^K = e + \delta  + 2k.
\end{equation}
Now using the relation between the semi-latus rectum and the semi-major axis $a^K$, $p^{K} = a^{K}(1-{e^{K}}^2)$, one obtains,
\begin{equation}
\label{apeq:aKaDD}
	a^K = a(1+\delta + 2k).
\end{equation}
Equating the radial coordinate $r=r^K$ where $r^K = a^K(1-e^K\cos E^K)$ is the radial coordinate of the osculating orbit and r is defined for the DD solution by equation \eqref{eq:DDdist}, we obtain at relevant order that $E^K(t=T_p) =0$ and therefore $T_p^K = T_P$. 
By further equating the directions, $\vr^{K}=\vr$, one then straightforwardly obtains $\omega^{K}=\omega$.

As mentioned in the text, the orbital inclination $i$ and the longitude of the line of ascending node $\Omega$ are not affected by the perturbations considered in this paper. In particular, it follows that the projected semi-major axis of the osculating orbit is immediately obtained by $x^{K} = a^K \sin i^K$.
\begin{table}
	\caption{Correspondence between the observable elements, the DD elements \citep{damour_general_1985}, and the elements of the osculating Keplerian orbit at periastron passage. \label{tab:elements}}
	\begin{tabular}{lll}
		Observable & DD & Keplerian  at $t=T_p$	\\
		\hline
		$\bar{x}$ & $x=\bar{x}$ & $x^K =\bar{x}(1+\delta + 2k) $ \\
		$\bar{n}$ & $n = \bar{n}(1-k)$ & $n^K = \bar{n}(1-2\delta - 4k)$ \\
		$\bar{\ks}$ & $\ks = \bar{\ks}$ & $\ks^K =\bar{\ks} + (\delta + 2k)\sin\bar{\omega} $ \\
		$\bar{\kc}$ & $\kc = \bar{\kc}$ & $\kc^K =\bar{\kc} + (\delta + 2k)\cos\bar{\omega} $ \\
		$\bar{T}_a$ & $T_a = \bar{T}_a + 2\ks/\bar{n}$ & $T_a^K  =  \bar{T}_a + 2\bar{\ks}/\bar{n}$ \\
		& & $ - 2\frac{\delta + 2k}{\bar{n}}(\bar{\omega} - \sin\bar{\omega} )$ \\
		\hline
		\multicolumn{3}{c}{Derived elements} \\
		\hline
		$\bar{\omega} = \arctan\frac{\bar{\ks}}{\bar{\kc}}$ & $\omega = \bar{\omega}$ & $\omega^K = \bar{\omega}$ \\
		$\bar{e} = \left(\bar{\kc}^2 + \bar{\ks}^2\right)^{1/2}$ & $e = \bar{e}$ & $e^K = \bar{e} + \delta + 2k$ \\
		$\bar{T}_p=\bar{T}_a + \frac{\bar{\omega}}{n}$ & $T_p = \bar{T}_p $ & $ T_p^K = \bar{T}_p$
	\end{tabular}
\end{table}

\section{Pseudo-DD solutions}
\label{secap:pseudoDD}
Here we show that, to first order in eccentricity and perturbation, and including secular terms (i.e. precession) the DD formalism can be formally extended to include any perturbation deriving from a potential of the form 
\newcommand{\Jp}{J}
\begin{equation}
\label{apeq:phip}
\Phi_\alpha(r) = - \Jp \frac{GM}{r} \frac{a^\alpha}{r^\alpha},
\end{equation}
where $\Jp \ll 1$ is the constant dimensionless magnitude of the perturbation, and $\alpha$ the perturbation index. 

The particular case $\alpha = 3$ corresponds to the spin-induced quadrupole moment when the spin axis is orthogonal to the orbital plane. This case can be directly solved by a conchoidal transformation of the equations of motion as shown in \citep{wex_timing_1998,damour_general_1985} and Section \ref{sec:DD}.

The same mathematical trick is however not generally applicable to other values of $\alpha$, but the mathematical simplicity and closed-form of the DD solutions makes it interesting to turn these solutions in a similar form if possible. In this paper, we are particularly interested by the case $\alpha=5$ which corresponds to the potential of an equilibrium tide.

We derive the variation of the orbital elements by inserting the potential $\Phi_p = -\Phi_\alpha$ into Lagrange's planetary equations \eqref{eq:llp}-\eqref{eq:llT0}, also using equations \eqref{eq:kc}-\eqref{eq:ks} and \eqref{eq:Ta}. Neglecting periodic terms of order $\sim\Jp e$, we obtain the following variations,
\begin{eqnarray}
\label{apeq:dkc}
\Delta \kc & = & \alpha\Jp\left[ \frac{1}{2}\left(1 - \alpha\right) n\ks c_0 - n s_1\right],\\
\Delta \ks & = & \alpha\Jp\left[ -\frac{1}{2}\left(1 - \alpha\right) n\kc c_0 + n c_1\right],\\
\Delta T_a & = & \alpha\Jp \left[\left(-2 + \kc(\alpha-1) \right)c_0 + \right. \nonumber \\
& & \left. \left(2 - 3\ks n \Delta t\right)c_1  + 3n\kc\Delta t s_1\right] ,
\label{apeq:dTa}
\end{eqnarray}
and $\Delta p  =\Delta i = \Delta \Omega = 0$. Using the definition of the mean motion $n=\sqrt{\frac{GM}{a^3}}$ and of the semi-latus rectum $p= a(1-e^2)$ we also obtain,
\begin{eqnarray}
\Delta a & = & 2\alpha \Jp a n \left(\ks c1 - \kc s1 \right),\\
\Delta n & = & 3\alpha\Jp n^2 \left(\kc s_1 - \ks c_1\right) .
\end{eqnarray}
We want the osculating orbital elements to match the DD orbital elements at the time of periastron passage $T_p$ (see appendix \ref{secap:osculating}). Therefore we integrate the variations from $T_p = T_a + \omega$ (to relevant order) to $t$, leading to:   $\Delta t = t  -T_a$, $c_0 = \Delta t - \omega/n$, $c1 = (\sin n\Delta t - \sin\omega)/n$ and $s_1 = -(\cos n\Delta t - \cos\omega)/n$. 

Inserting the perturbed orbital elements (see equation \eqref{eq:elementperturbed}) in the exact 2-body solution (equation \eqref{eq:DDpos}-\eqref{eq:DD3rdlaw} with $k=\delta=\delta_v=\delta_r=0$) one derives the position vector of the perturbed motion, equation \eqref{eq:DDpos}. One then expands the result to first order in eccentricity and perturbation while retaining secular effects using equations \eqref{eq:Ee2} and \eqref{eq:ve2}, that is keeping terms of order $\sim e, \sim\Jp $ and to any order in $\sim te\Jp$. 

At this stage the solution is expressed in terms of Keplerian orbital elements (superscript K in appendix \ref{secap:osculating}). We thus need to replace them by the ``pseudo-DD''elements using in particular  equations \eqref{apeq:eKeDD} and \eqref{apeq:aKaDD} to replace $e^K$ and $a^K$. One then finds, provided that  
\begin{eqnarray}
\label{apeq:pseudoDDkdelta}
k = -\delta = \frac{\alpha(\alpha - 1)}{2}\Jp,
\end{eqnarray}
that the perturbed position vector exactly matches the position vector derived from the DD solution with $k$ and $\delta$ defined as in equation \eqref{apeq:pseudoDDkdelta}. Explicitly, the position vector reads
\begin{equation}
\label{apeq:perturbtraj}
\vr = a
\left(
\begin{matrix}
\cos\sigma  - \frac{3}{2}\kc' + \frac{1}{2}\ks'\sin 2\sigma + \frac{1}{2}\kc'\cos 2\sigma \\
\sin\sigma  - \frac{3}{2}\ks' + \frac{1}{2}\kc'\sin 2\sigma - \frac{1}{2}\ks'\cos 2\sigma \\
0
\end{matrix}\right)_{(\vx,\vy,\vz)}
\end{equation}
with $\sigma = (1-k)n(t-T_a) - 2\ks = \bar{n}(t-\bar{T}_a)$, $\kc' = e\cos\left(\omega + k\sigma\right)$ and $\ks' = e\sin\left(\omega + k\sigma\right)$ are defined as in \eqref{eq:kcp} and \eqref{eq:ksp}, $(\vx,\vy,\vz)$ is defined in Section \ref{sec:DD}.

%


\bsp	
\label{lastpage}
\end{document}